\documentclass[fleqn,usenatbib]{mnras}

\usepackage{graphicx}	
\usepackage{amsmath}	
\usepackage{amssymb}	
\usepackage{hyperref}
\usepackage{upgreek}
\usepackage{longtable}
\usepackage{supertabular} 
\usepackage{booktabs}
\usepackage{placeins}
\usepackage{lscape}
\usepackage{xspace} 
\usepackage{float}
\usepackage{siunitx}
\usepackage{tablefootnote}
\usepackage{threeparttable}
\usepackage{pdflscape}
\usepackage{enumitem}

\graphicspath{{./Figures/}}

\newcommand{\etacar}{$\eta$~Car\xspace}
\newcommand{\jbu}{AT~2016jbu\xspace}
\newcommand{\ip}{SN~2009ip\xspace}
\newcommand{\bh}{SN~2015bh\xspace}
\newcommand{\gc}{SN~2013gc\xspace}
\newcommand{\bdu}{SN~2016bdu\xspace}

\newcommand{\lsq}{LSQ13zm\xspace}
\newcommand{\al}{SN~1996al\xspace}
\newcommand{\cnf}{SN~2018cnf\xspace}

\newcommand{\apt}{{\sc AutoPhOT \xspace}}

\newcommand{\msun}{M$_\odot$\xspace}
\newcommand{\kms}{km~s$^{-1}$\xspace}

\DeclareSIUnit\angstrom{\text {Å}}
\newcommand{\orcid}[1]{\href{https://orcid.org/#1}{\includegraphics[width=10pt]{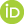}}}

\newcommand{\paperII}{\textcolor{blue}{Paper II}\xspace}

\title[Photometric and spectroscopic evolution of the interacting transient, \jbu]{Photometric and spectroscopic evolution of the interacting transient, \jbu (Gaia16cfr)}

\author[S. J. Brennan et al.]{S. J. Brennan$^{1}$\orcid{0000-0003-1325-6235}%
\thanks{Contact e-mail: \href{sean.brennan2@ucdconnect.ie}{sean.brennan2@ucdconnect.ie}},
M. Fraser$^{1}$\orcid{0000-0003-2191-1674} ,
J. Johansson$^{2}$\orcid{0000-0001-5975-290X},
A. Pastorello$^{3}$\orcid{0000-0002-7259-4624},
R. Kotak$^{4}$
H. F. Stevance$^{5}$,\newauthor 
T. -W. Chen$^{6,7}$\orcid{0000-0002-1066-6098},
J. J. Eldridge$^{5}$\orcid{0000-0002-1722-6343},
S. Bose$^{8,9}$,
P. J. Brown$^{10}$\orcid{0000-0001-6272-5507},
E. Callis$^{1}$\orcid{0000-0002-1178-2859},
R. Cartier$^{11}$,\newauthor 
M. Dennefeld$^{12}$,
Subo Dong$^{13}$,
P. Duffy$^{1}$,
N. Elias-Rosa$^{14,15}$\orcid{0000-0002-1381-9125},
G. Hosseinzadeh$^{16}$\orcid{0000-0002-0832-2974},\newauthor 
E. Hsiao$^{17}$\orcid{0000-0003-1039-2928},
H. Kuncarayakti$^{18,19}$,
A. Martin-Carrillo$^{1}$, 
B. Monard$^{20}$,
A. Nyholm$^{6}$,
G. Pignata$^{21,22}$\orcid{0000-0003-0006-0188},\newauthor
D. Sand$^{23}$\orcid{0000-0003-4102-380X},
B. J. Shappee$^{24}$\orcid{0000-0003-4631-1149},
S. J. Smartt$^{25}$,
B. E. Tucker$^{26,27,28}$\orcid{0000-0002-4283-5159},
L. Wyrzykowski$^{29}$\orcid{0000-0002-9658-6151},\newauthor 
H. Abbot$^{26}$,
S. Benetti$^{3}$\orcid{0000-0002-3256-0016},
J. Bento$^{26}$,
S. Blondin$^{30,31}$\orcid{0000-0002-9388-2932},
Ping Chen$^{28}$,
A. Delgado$^{32,33}$,
L. Galbany$^{34}$\orcid{0000-0002-1296-6887},\newauthor 
M. Gromadzki$^{29}$\orcid{0000-0002-1650-1518},
C. P. Guti\'errez$^{19,35}$\orcid{0000-0003-2375-2064},
L. Hanlon$^{1}$,
D. L. Harrison$^{32,36}$\orcid{0000-0001-8687-6588},
D. Hiramatsu$^{37,38,54,55}$\orcid{0000-0002-1125-9187},\newauthor 
S. T. Hodgkin$^{32}$\orcid{0000-0002-5470-3962},
T. W.-S. Holoien$^{39}$\orcid{000-0001-9206-3460},
D. A. Howell$^{37,38}$\orcid{0000-0003-4253-656X},
C. Inserra$^{40}$\orcid{0000-0002-3968-4409},
E. Kankare$^{4}$\orcid{0000-0001-8257-3512},\newauthor 
S. Koz{\l}owski$^{29}$\orcid{0000-0003-4084-880X},
T. E. M\"{u}ller-Bravo$^{41,53}$\orcid{0000-0003-3939-7167},
K. Maguire$^{42}$\orcid{0000-0002-9770-3508},
C. McCully$^{37,38}$\orcid{0000-0001-5807-7893},
P. Meintjes$^{43}$,\newauthor 
N. Morrell$^{44}$\orcid{0000-0003-2535-3091},
M. Nicholl$^{45,46}$,
D. O'Neill$^{25}$,
P. Pietrukowicz$^{29}$\orcid{0000-0002-2339-5899},
R. Poleski$^{29}$\orcid{0000-0002-9245-6368},
J. L. Prieto$^{22,47}$,\newauthor 
A. Rau$^{7}$,
D. E. Reichart$^{48}$\orcid{0000-0002-5060-3673},
T. Schweyer$^{6,7}$,
M. Shahbandeh$^{49}$,
J. Skowron$^{29}$\orcid{0000-0002-2335-1730},\newauthor 
J. Sollerman$^{6}$\orcid{0000-0003-1546-6615},
I. Soszy{\'n}ski$^{29}$\orcid{ 0000-0002-7777-0842},
M. D. Stritzinger$^{50}$\orcid{0000-0002-5571-1833},
M. Szyma{\'n}ski$^{29}$\orcid{ 0000-0002-0548-8995},
L. Tartaglia$^{3}$\orcid{0000-0003-3433-1492},\newauthor 
A. Udalski$^{29}$\orcid{0000-0001-5207-5619},
K. Ulaczyk$^{29,51}$\orcid{0000-0001-6364-408X},
D. R. Young$^{52}$\orcid{0000-0002-1229-2499},
M. van Leeuwen$^{32}$,
B. van Soelen$^{43}$
\\
The authors' affiliations are shown in Appendix \ref{app:affiliations}.
}

\pubyear{2021}

\begin{document}


\label{firstpage}
\pagerange{\pageref{firstpage}--\pageref{lastpage}}
\maketitle

\begin{abstract}

We present the results from a high cadence, multi-wavelength observation campaign of \jbu, (aka Gaia16cfr) an interacting transient. This dataset complements the current literature by adding higher cadence as well as extended coverage of the lightcurve evolution and late-time spectroscopic evolution. Photometric coverage reveals that \jbu underwent significant photometric variability followed by two luminous events, the latter of which reached an absolute magnitude of $M_V\sim-18.5$~mag. This is similar to the transient \ip whose nature is still debated. Spectra are dominated by narrow emission lines and show a blue continuum during the peak of the second event. \jbu shows signatures of a complex, non-homogeneous circumstellar material (CSM). We see slowly evolving asymmetric hydrogen line profiles, with velocities of 500~\kms seen in narrow emission features from a slow moving CSM, and up to 10,000~\kms seen in broad absorption from some high velocity material. Late-time spectra ($\sim$~+1 year) show a lack of forbidden emission lines expected from a core-collapse supernova  and are dominated by strong emission from H, \ion{He}{I} and \ion{Ca}{II}. Strong asymmetric emission features, a bumpy lightcurve, and continually evolving spectra suggest an inhibit nebular phase. We compare the evolution of H$\alpha$ among \ip-like transients and find possible evidence for orientation angle effects. The light-curve evolution of \jbu suggests similar, but not identical, circumstellar environments to other \ip-like transients.
\end{abstract}




\defcitealias{Kilpatrick2018}{K18}


\section{Introduction}\label{sec:intro}

Massive stars that eventually undergo core-collapse when surrounded by some dense circumstellar material (CSM) are known as Type IIn supernovae (SNe) \citep{Schlegel90, Filippenko1997, Fraser20}. This is signified in spectra by a bright, blue continuum with narrow H and $\ion{He}{I}$ emission lines at early times. Type IIn SNe spectra show narrow ($\sim 100-500$~\kms) components arising in the photo-ionised, slow moving CSM. Intermediate width emission lines ($\sim 1000$~\kms) arise from either electron scattering of photons in narrower lines or emission from gas shocked by supernova (SN) ejecta. Some events also show very broad emission or absorption features ($\sim 10,000$~\kms) arising from fast ejecta, typically associated with material ejected in a core-collapse explosion.

The existence of the dense CSM indicates that the Type IIn progenitors have high mass-loss rates shortly before their terminal explosion. This dense material at the end of a star's life can come from several pathways \citep[see reviews by][for further detail.]{Puls_2008,Smith2014b, Fraser20}.

Complicating this picture are a growing number of extragalactic transients that show narrow emission lines in their spectra (indicating CSM) but have much fainter absolute magnitudes than most typical Type IIn SNe. These events are often termed \textit{SN Impostors} \citep{VanDyk00, Maund_2006,PastoFraser19}, and are believed in many cases to be extra-galactic Luminous Blue Variables (LBVs) experiencing giant eruptions \citep[e.g. SN~2000ch;][]{Wagner_2004,Pastorello_2010}. These eruptions do not completely destroy their progenitors.

Perhaps the best studied exemplar of the confusion between LBVs, SN impostors, and genuine Type IIn SNe is \ip. \ip was found on 2009 August 26 at $\sim$~17.9 mag in NGC 7259 by CHASE project team members \citep{maza_2009}. This transient was originally classified as a Type IIn SN, and then re-classed as an impostor when it became clear that the progenitor had survived. \ip was characterized by a years-long phase of erratic variability that ended with two luminous outbursts a few weeks apart in 2012 \citep{Li09,drake10,Margutti12,Pastorello2013,Fraser13,Smith2014,Graham2014}.

From pre-explosion images taken 10 years prior to the 2009 discovery, the progenitor star of \ip was suggested to be a LBV with a mass of 50--80~\msun\ \citep{smith10,Foley2011}. There is much debate on the fate of \ip. Some argue that \ip has finally exploded as a genuine Type IIn SN during the 2012 outburst \citep{prieto13,Mauerhan2013}. However, other authors remain agnostic as to \ip's fate as a CCSN, pointing to the absence of any evidence for nucleosynthesised material in late-time spectra, as well as \ip not fading significantly below the progenitor magnitude \citep{Fraser13,Margutti2014,Fraser2015}. Since the discovery of \ip, a number of remarkably similar transients have been found. The growing family of \ip-like transients share similar spectral and photometric evolution. \ip-like transients have the following observable traits.

\begin{enumerate}[label=\bfseries \arabic*:,leftmargin=*,labelindent=1em]

 \item History of variability lasting (at least) $\sim$~10 years with outbursts reaching $M_r\sim-11\pm3$~mag.
 
 \item Two bright luminous events with the first peak reaching a magnitude of $M_r\sim-13\pm2$~mag followed by the second peak reaching $M_r\sim-18\pm1$~mag several weeks later.
 
 \item Spectroscopically similar to a Type IIn SN i.e. narrow emission features and a blue continuum at early times.

 \item Restrictive upper limits to the mass of any explosively synthesised $^{56}$Ni.
 
\end{enumerate}

\begin{figure}
\centering
\includegraphics[width= \columnwidth]{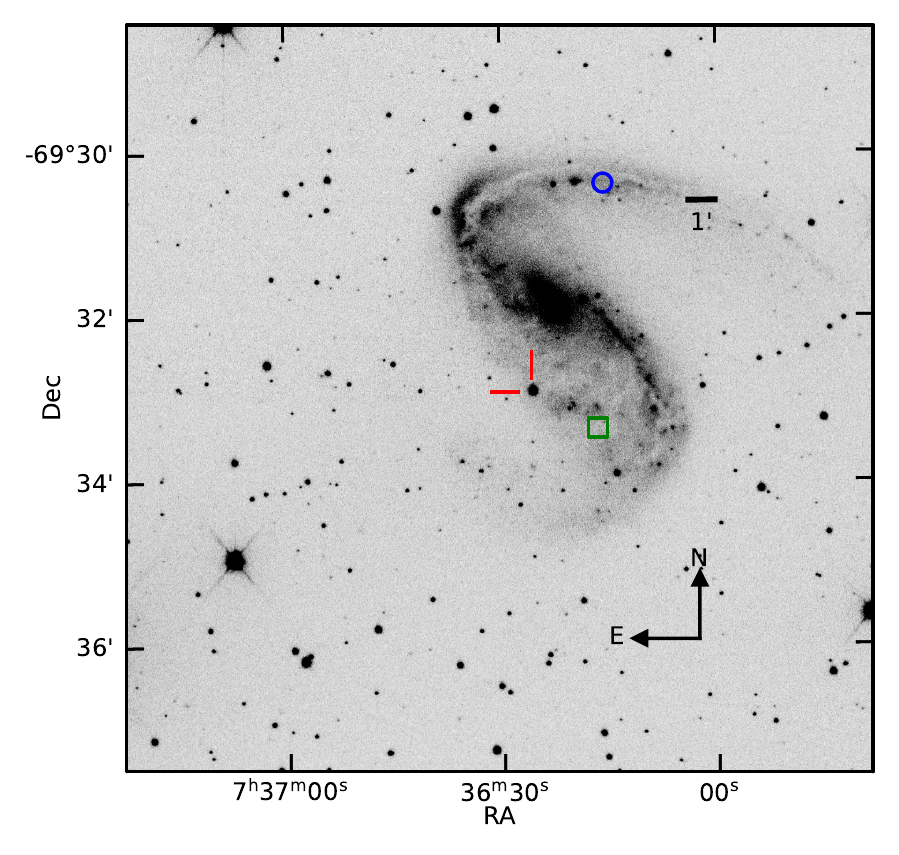}
\caption{Finder chart for \jbu. Image is a 60s \textit{\textit{r}-band} exposure taken with the LCO 1-m. \jbu is situated to the south-east of the spiral galaxy NGC~2442 nucleus and is indicated with a red cross reticle in the center of the image. This location lies on the outskirts of a \textit{Superbubble} \citep{Pancoast_2010}, with a high star formation rate. We also include the location of the Type Ia SN~2015F (blue circle, north west of image center; \citealp{Cartier17}) and the Type II SN~1999ga (green square, south west of image center; \citealp{Pastorello2009}).}
\label{fig:finder}
\end{figure}

In this paper we focus on one such \ip-like transient. \jbu (also known as Gaia16cfr; \citealp{Bose17}) was discovered at RA. = 07:36:25.96, DEC. = $-$69:32:55.25 (J2000) by the {\it Gaia} satellite on 2016 December 1 with a magnitude of $G=$19.63 (corresponding to an absolute magnitude of $-11.97$~mag for our adopted distance modulus). The Public ESO Spectroscopic Survey for Transient Objects (PESSTO) collaboration \citep{Smartt15} classified \jbu as an \ip-like transient due to its spectral appearance and apparent slow rise \citep{atel9938}. \citet{atel9938} also finds that the progenitor of \jbu seen in archival Hubble Space Telescope ({\it HST}) images is consistent with a massive ($<$~30~\msun) progenitor. The transient was independently discovered by B. Monard in late December who reported the likely association of \jbu to its host, NGC~2442. \jbu is situated to the south of NGC~2442, a spiral galaxy commonly referred to as the \textit{Meathook Galaxy}. NGC~2442 has hosted two other SNe including SN~1999ga, a low luminosity Type II SN \citep{Pastorello2009} and SN~2015F, a Type Ia SN \citep{Cartier17}. We mark their respective locations in Fig.~\ref{fig:finder}. \citet{Bose17} and \citet{ATel11726} reported initial spectroscopic observations and classification of \jbu.

\jbu has been previously studied by \citet{Kilpatrick2018} (hereafter referred to as \citetalias{Kilpatrick2018}). \citetalias{Kilpatrick2018} finds that \jbu appears similar to a Type IIn SN, with narrow emission lines and a blue continuum. The {\it Gaia} lightcurve shows that \jbu has a double-peaked lightcurve showing two distinct events (we refer to these events as \textit{Event A} and \textit{Event B}). This is common in \ip-like transient with \textit{Event B} reaching an absolute magnitude of r$\sim~-18$~mag. H$\alpha$ displays a double-peaked profile a few weeks after maximum brightness, indicating a complex CSM environment. \citetalias{Kilpatrick2018} model H$\alpha$ using a multi-component line profile including a shifted blue emission feature that grows with time, with their final profile similar to that of the Type IIn \bh \citep{Elisa-Rosa2016,Thone2017} at late times.

Using {\it HST} images, spanning 10 years prior to the 2016 transient, \citetalias{Kilpatrick2018} reports that \jbu underwent a series of outbursts in the decade prior, similar to \ip. and finds the progenitor is consistent with a $\sim$~18~\msun progenitor star, with strong evidence of reddening by circumstellar (CS) dust (which would allow for a higher mass). Performing dust modelling using Spitzer photometry, \citetalias{Kilpatrick2018} find the spectral energy distribution (SED) $\sim$~10 years prior is fitted well with a warm dust shell at 120~AU. They find that, given typical CSM velocities, it is unlikely that this dusty shell is in the immediate vicinity of the progenitor and is unlikely to be seen during the 2016 event. This means that the progenitor of \jbu was experiencing episodic mass loss within years to decades of its most recent explosion.

This paper focuses on photometry and spectra obtained for \jbu which is not covered by \citetalias{Kilpatrick2018}. In particular, this includes searching through historic observations of \jbu's host, NGC~2442 for signs of variability, as is expected for \ip-like transients, as well as presenting high cadence data for \textit{Event A} and the late time photometric and spectroscopic evolution.

We take the distance modulus for NGC~2442 to be $31.60\pm0.06$~mag, which is a weighted average of the values determined from {\it HST} observations of Cepheids \citep[$\upmu=31.511\pm0.053~mag~$;][]{Riess16} and from the SN Ia 2015F \citep[$\upmu=31.64\pm0.14$~mag;][]{Cartier17}. This corresponds to a metric distance of $20.9\pm0.58$~Mpc. We adopt a redshift of z=0.00489 from H I Parkes All Sky Survey \citep{Wong2006}. The foreground extinction towards NGC~2442 is taken to be $A_V=0.556$~mag, from \citet{Schlafly11} via the NASA Extragalactic Database (NED;\footnote{\url{https://ned.ipac.caltech.edu/}}). We correct for foreground extinction using $R_V=3.1$ and the extinction law given by \citealp{Cardelli1989}. We do not correct for any possible host galaxy or circumstellar extinction, however we note that the blue colors seen in the spectra of \jbu do not point towards significant reddening by dust. We take the {\it V}-band maximum during the second, more luminous event in the lightcurve (as determined through a polynomial fit) as our reference epoch (MJD $57784.4\pm0.5$; 2017 Jan 31).

This is the first of two papers discussing \jbu. In this paper (Paper I), we report spectroscopic and photometric observations of \jbu. In Sect.~\ref{sec:obsdata} we present details of data reduction and calibration. In Sect.~\ref{sec:phot_evo} and Sect.~\ref{sec:spec} we discuss the photometric and spectroscopic evolution of \jbu respectively. In Sect.~\ref{sec:dicussion} we compare \jbu to \ip-like transients, and also consider the observational evidence for core-collapse.

In \citealt{brennan2021b} (hereafter \paperII)  focuses on the progenitor of \jbu, its environment and using modelling to constrain the physical properties of this event.

\section{Observational data}\label{sec:obsdata}

The optical lightcurve evolution of \jbu has been previously discussed in \citetalias{Kilpatrick2018}. Their analysis covers \textit{Event B} up to $\sim$+140~days past maximum brightness. We present a higher cadence photometric dataset that covers both \textit{Event A}, \textit{Event B}, as well as late-time observations up to $\sim$+575~d. This high cadence dataset allows for a more detailed photometric analysis of \jbu which will be discussed in Sect.\ref{sec:dicussion}.
\citetalias{Kilpatrick2018} discuss the spectral evolution of \jbu from $-$27~days until +118~days. Our observational campaign presented here continues contains increased converge during this period as well as observations up until $+420$~days allowing for late time spectral followup.

\subsection{Optical imaging and reduction}\label{sec:image_reduction}
Optical imaging of \jbu in {\it BVRri} filters was obtained with the 3.58m ESO New Technology Telescope (NTT) + EFOSC2, as part of the ePESSTO survey. All images were reduced in the standard fashion using the PESSTO pipeline \citep{Smartt15}; in brief images were bias and overscan subtracted, flat fielded, before being cleaned of cosmic rays using a Laplacian filter \citep{lacosmic}. Further optical imaging was obtained from the Las Cumbres Observatory network of robotic 1-m telescopes as part of the  Global  Supernova  Project. These data were reduced automatically by the {\sc banzai} pipeline, which runs on all Las Cumbres Observatory (LCO) Global Telescope  images \citep{Brown2013}. Images were also obtained from the Watcher telescope. Watcher is a 40 cm robotic telescope located at Boyden Observatory in South Africa \citep{Fren04}. It is equipped with an Andor IXon EMCCD camera providing a field of view of 8$\times$8 arcmin. The Watcher data were reduced using a custom made pipeline written in {\sc Python}. 

\jbu was monitored using the Gamma-Ray Burst Optical/Near-Infrared Detector (GROND; \citet{Grei08}), a 7-channel imager that collects multi-color photometry simultaneously with Sloan-\textit{griz} and {\it JHK/Ks} bands, mounted at the 2.2 m MPG telescope at ESO La Silla Observatory in Chile. The images were reduced with the GROND pipeline \citep{Kruhler2008}, which applies de-bias and flat-field corrections, stacks images and provides astrometry calibration. Due to the bright host galaxy we disabled line by line fitting of the sky subtraction for the GROND NIR data since this caused over subtraction artifacts. Since the photometry background estimation is limited by the extended structure of the host galaxy and not the large-scale variation in the background of the image. We do not expect any adverse effects from this change. 

Unfiltered imaging of \jbu was also obtained by B. Monard. Observations of \jbu were taken at the Kleinkaroo Observatory (KKO), Calitzdorp (Western Cape, South Africa) using a 30cm telescope Meade RCX400 f/8 and CCD camera SBIG ST8-XME in 2$\times$2 binned mode. Unfiltered images were taken with 30s exposures, dark subtracted and flat fielded and calibrated against \textit{r}-band sequence stars. Nightly images resulted from stacking (typically 5 to 8) individual images. 

We also recovered a number of archival images covering the site of \jbu. Two epochs of {\it g} and {\it r} imaging from the Dark Energy Camera (DECam) \citep{Flaugher_2015} mounted on the 4~m Blanco Telescope at the Cerro Tololo Inter-American Observatory (CTIO) were obtained from the NOIRLab Astro Data Archive. The science-ready reduced ``InstCal'' images were used in our analysis. In addition to these, we downloaded deep imaging taken in 2005 with the MOSAIC-II imager (the previous camera on the 4~m Blanco Telescope). As for the DECam data, the ``InstCal'' reductions of MOSAIC-II images were used. We note that the filters used for the MOSAIC-II images (Harris $V$ and $R$, Washington $C$ \citealp{Harris1979}) are different from the rest of our archival dataset. The Harris filters were calibrated to Johnson-Cousins $V$ and $R$. The Washington $C$ filter data is more problematic, as this bandpass lies between Johnson-Cousins $U$ and $B$. We calibrated our photometry to the latter, but this should be interpreted with appropriate caution.

Deep Very Large Telescope (VLT) + OmegaCAM images taken with {\it i}, {\it g}, and {\it r} filters in 2013, 2014, and 2015, respectively, were downloaded from the ESO archive. The Wide Field Imager (WFI) mounted on the 2.2-m MPG telescope at La Silla also observed NGC~2442 on a number of occasions between 1999 and 2010 in {\it B}, {\it V}, and {\it R}; these images are of particular interest as they are quite deep, and extend our monitoring of the progenitor as far back as $-15$ years. Both the OmegaCAM and WFI data were reduced using standard procedures in \textsc{IRAF}\footnote{\textsc{IRAF}is distributed by the National Optical Astronomy Observatory, which is operated by the Association of Universities for Research in Astronomy (AURA) under cooperative agreement with the National Science Foundation}.

NED contains a number of historical images of NGC~2442, dating back to 1978. We examined each of these but found none that contained a credible source at the position of \jbu.

Several transient surveys also provided photometric measurements for \jbu. {\it Gaia} {\it G}-band photometry for \jbu was downloaded from the Gaia Science Alerts web pages. As this photometry was taken with a broad filter that covers approximately {\it V} and {\it R}, we did not attempt to calibrate it onto the standard system. {\it V}-band imaging was also taken as part of the All-Sky Automated Survey for Supernovae \citep[ASAS-SN][]{Shap14,Koch17}\footnote{\url{http://www.astronomy.ohio-state.edu/asassn/index.shtml}}.

The OGLE IV Transient Detection System \citep{2013AcA....63....1K, 2014AcA....64..197W} also identified \jbu, and reported $I$-band photometry via the OGLE webpages\footnote{\url{http://ogle.astrouw.edu.pl/ogle4/transients/}}. 

The Panchromatic Robotic Optical Monitoring and Polarimetry Telescopes (PROMPT) \citep{Reichart2005} obtained imaging of \jbu in $BVRI$ filters; and as discussed in Sect.~\ref{sec:phot_comp}, unfiltered PROMPT observations of NGC~2442 were also used to constrain the activity of the progenitor of \jbu over the preceding decade. Images were taken with the PROMPT1, PROMPT3, PROMPT4, PROMPT6, PROMPT7 and PROMPT8 robotic telescopes (all located at the CTIO). PROMPT4 and PROMPT6 have a diameter of 40~cm while PROMPT1, PROMPT3 and PROMPT8 have a diameter of 60~cm and PROMPT7 has a diameter of~80 cm. All images collected with the PROMPT units were dark subtracted and flat-field corrected. In case multiple images were taken in consecutive exposures, the frames were registered and stacked to produce a single image.

NGC~2442 was also serendipitously observed with the FOcal Reducer/low dispersion Spectrograph 2 (FORS2) as part of the late-time follow-up campaign for SN~2015F \citep{Cartier17}. Unfortunately, most of these data were taken with relatively long exposures, and \jbu was saturated. However, a number of pre-discovery images from the second half of 2016, as well as late time images from 2018 are of use. These data were reduced (bias subtraction and flat fielding) using standard {\sc iraf} tasks.

\subsection{UV Imaging}\label{sec:uv}

UV and optical imaging was obtained with the {\it Neil Gehrels Swift Observatory} ({\it Swift}) with the Ultra-Violet Optical Telescope (UVOT). The pipeline reduced data was downloaded from the {\it Swift} Data Center. The photometric reduction follows the same basic outline as \citet{Brown_2009}. In short, a 5\arcsec\ radius aperture is used to measure the counts for the coincidence loss correction, a 3\arcsec\ source aperture (based on the error) was used for the aperture photometry and applying an aperture correction as appropriate (based on the average PSF in the {\it Swift} HEASARC's calibration database (CALDB) and zeropoints from \citet{Breeveld2011}. 

Subsequent to the photometric reduction of our \textit{Swift} data, there was an update to the \textit{Swift} CALDB with time dependant zero-points which we have not accounted for. Given that our \textit{Swift} observations occurred in early 2017, this would amount to a $\sim~3\%$ shift in zero-point and would not lead to a significant change in our lightcurve.

\subsection{NIR imaging}\label{sec:NIR}

Near-infrared imaging was obtained with NTT+SOFI as part of the ePESSTO survey, and with GROND as mentioned previously. In both cases {\it JHK/Ks} filters were used. SOFI data were reduced using the PESSTO pipeline \citep{Smartt15}. Data were corrected for flat-field and illumination, sky subtraction was performed using (in most instances) off-target dithers, before individual frames were co-added to make a science-ready image.

In addition to the follow-up data obtained for \jbu with SOFI, we examined pre-discovery SOFI images taken as part of the PESSTO follow-up campaign for SN~2015F. We downloaded reduced images from the ESO Phase 3 archive which covered the period up to April 2014. Two subsequent epochs of SOFI imaging from 2016 Oct were taken after PESSTO SSDR3 was released, and so we downloaded the raw data from the ESO archive, and reduced these using the PESSTO pipeline as for the rest of the SOFI follow-up imaging.

Fortuitously, the ESO VISTA telescope equipped with VIRCAM observed NGC~2442 as part of the VISTA Hemisphere Survey (VHS) in Dec. 2016. We downloaded the reduced images as part of the ESO Phase 3 data release from VHS via ESO Science Portal. Photometry was performed using \apt, see Sect.~\ref{sec:autophot}.

\subsection{MIR imaging}

We queried the WISE data archive at the NASA/IPAC infrared science archive, and found that \jbu was observed in the course of the NEOWISE reactivation mission \citep{Neowise}. As the spatial resolution of {\it WISE} is low compared to our other imaging, we were careful to select only sources that were spatially coincident with the position of \jbu. There were numerous detections of \jbu in the $W1$ and $W2$ bands over a one week period shortly before the maximum of \textit{Event B} (MJD $57784.4\pm0.5$). The profile-fitted magnitudes measured for each single exposure (L1b frames) were averaged within a 1 day window.

We also examined the pre-explosion images covering the site of \jbu in the Spitzer archive, taken on 2003 Nov. 21 (MJD 52964.1). Some faint and apparently spatially extended flux can be seen at the location of \jbu in $Ch1$, although there is a more point-like source present in $Ch2$. No point source is seen in Ch3 and Ch4. \citetalias{Kilpatrick2018} report values of $0.0111\pm0.0032$~mJy and $0.0117\pm0.0027$~mJy in $Ch1$ and $Ch2$ (corresponding to magnitude of 18.61~mag and 17.917~mag respectively) and similarly do not detect a source in $Ch4$ and $Ch4$ for the 2003 images.

\subsection{X-ray Imaging}\label{sec:x_ray_obs}

A target of opportunity observation (ObsID: 0794580101) was obtained with XMM-Newton \citep{Jans01} on 2017 Jan 26 (MJD 57779) for a duration of $\sim~57$~ks. The data from EPIC-PN \citep{Stru01} were analysed using the latest version of the Science Analysis Software, SASv18\footnote{\url{http://xmm.esac.esa.int/sas/}} including the most updated calibration files. The source and background were extracted from a 15\arcsec\ region avoiding a bright nearby source. Standard filtering and screening criteria were then applied to create the final products.

X-ray imaging was also taken with the XRT on board {\it Swift}. These observations are much less sensitive than the XMM-Newton data, and so we do not expect a detection. Using the online XRT analysis tools\footnote{\url{https://www.swift.ac.uk/user_objects/}} \citep{Evans07,Evans09} we co-added all XRT images covering the site of \jbu\ available in the Swift data archive. No source was detected coincident with \jbu in the resulting $\sim$~100~ks stacked image. 

\subsection{Photometry with the {\sc AutoPHoT} pipeline}\label{sec:autophot}

The dataset presented in this paper for \jbu\ comprises approximately $\sim$~3000 separate images from around 20 different telescopes. To expedite photometry on such large and hetrogeneous datasets, we have developed a new photometric pipeline called \apt ({\sc AUTOmated PHotometry Of Transients};\cite{autophot}). \apt has been used to measure all photometry presented in this paper, with the exception of imaging from space telescopes (i.e. {\it Swift}, {\it Gaia}, {\it WISE}, {\it Spitzer}, {\it XMM-Newton OM} and {\it HST}), as well as from ground based surveys which have custom photometric pipelines (i.e. ASAS-SN and OGLE).

\apt\footnote{\url{https://github.com/Astro-Sean/autophot}}\footnote{\url{https://anaconda.org/astro-sean/autophot}} is a {\sc Python3}-based photometry pipeline built on a number of commonly used astronomy packages, mostly from {\sc astropy}. \apt is able to handle hetrogeneous data from different telescopes, and performs all steps necessary to produce a science-ready lightcurve with minimal user interaction.

\begin{figure*}
\centering
\includegraphics[width=\textwidth]{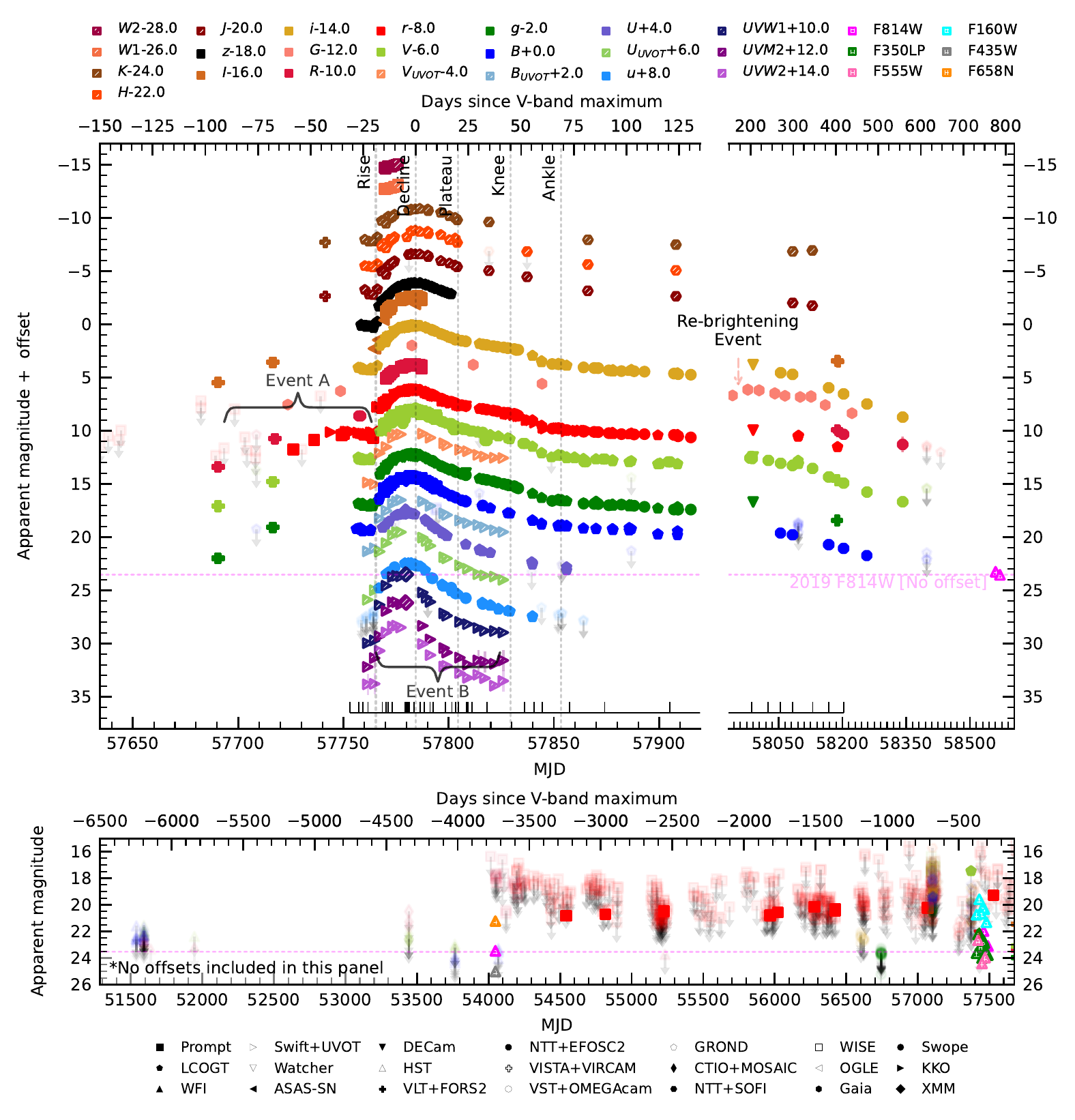}
\caption{The complete multi-band observed photometry for \jbu. The upper panel covers the period from the start of \textit{Event A} (First detection at $-91$~d from VLT+FORS2) until the end of our monitoring campaign $\sim$~2 years after \textit{Event B} peak. Offsets (listed in the legend) have been applied to each filter for clarity in the upper panels only. Note that there is a change in scale in the X-axis after 135 days. We indicate \textit{Event A} and the rise and decline of the peak of \textit{Event B}. Epochs where spectra were taken are marked with vertical ticks. We also include the published Swope photometry from \citetalias{Kilpatrick2018} (given as filled circles) to demonstrate that our photometry is consistent. We include a horizontal magenta dotted line in all panels to demonstrate the early 2019 $F814W$ magnitudes (\paperII). We only plot error bars greater than 0.1~mag. The lower panel shows detections and upper limits over a period from $\sim~18$ years prior to \textit{Event A}. No offsets are included in this panel; light points with arrows show upper limits, while solid points are detections. }
\label{fig:lc}
\end{figure*}

In brief, \apt will build a model for the Point Spread Function (PSF) in an image from bright isolated sources in the field (if no suitable sources are present then \apt will fall back to aperture photometry). This PSF is then fitted to the transient to measure the instrumental magnitude. To calibrate the instrumental magnitude onto the standard system (either AB magnitudes for Sloan-like filters or Vega magnitudes for Johnson-Cousins filters) for this work on \jbu, the zeropoint for each image is found from catalogued standards in the field. For {\it griz} filters, the zeropoint was calculated from magnitudes of sources in the field taken from the SkyMapper Southern Survey \citep{Onke19}. For Johnson-Cousins filters, we used the tertiary standards in NGC~2442 presented by \citet{Pastorello2009}. In the case of the NIR data ({\it JHK}) we used sources taken from the Two Micron All Sky Survey \citep[2MASS;][]{Skru06}. There is no \textit{u}-band photometry covering this portion of the sky. We use \textit{U}-band photometry from \citet{Cartier17} and convert to \textit{u}-band using Table 1 in \citet{Jester_2005}. We include Swope photometry from \citetalias{Kilpatrick2018} in Fig.~\ref{fig:lc} to show that our \textit{u}-band is consistent.

\apt utilises a local version of {\sc Astrometry.net}\footnote{\url{http://astrometry.net/}} \citep{Barron2008} for astrometric calibration when image astrometric calibration meta-data is missing or incorrect. In instances where \jbu could not be clearly detected in an image, \apt performs template subtraction using {\sc hotpants}\footnote{\url{https://github.com/acbecker/hotpants}} \citep{hotpants}, before doing forced photometry at the location of \jbu. Based on the results of this, we report either a magnitude or a $3\sigma$ upper limit to the magnitude of \jbu. Artificial sources of comparable magnitude were injected and recovered to confirm these measurements and to determine realistic uncertainties, accounting for the local background and the presence of additional correlated noise resulting from the template subtraction.

Finally, in order to remove cases where a poor subtraction leads to spurious detections, we require that the FWHM of any detected source agrees with the FWHM measured for the image to within one pixel, as well as being above our calculated limiting magnitude. In practice we find these are good acceptance tests to avoid false positives, especially in the pre-discovery lightcurve of \jbu.

We present the observed lightcurve of \jbu in Fig.~\ref{fig:lc}, and show a portion of the tables of calibrated photometry in Appendix \ref{app:appendix_phot} (the full tables are presented in the online supplementary materials).

\subsection{Spectroscopic Observations}

Most of our spectroscopic monitoring of \jbu was obtained with NTT+EFOSC2 through the ePESSTO collaboration. With the exception of the first classification spectrum reported by \citet{atel9938}, observations were taken with grisms Gr\#11 and Gr\#16, which cover the range of 3345--7470 \AA\ and 6000--9995 \AA\ at resolutions of R$\sim$~390 and R$\sim$~595, respectively. 

The EFOSC2 spectra were reduced using the PESSTO pipeline; in brief, two-dimensional spectra were trimmed, overscan and bias subtracted, and cleaned of cosmic rays. The spectra were flat-fielded using either lamp flats taken during daytime (Gr\#11), or that were taken immediately after each science observation in order to remove fringing (in the case of Gr\#16). An initial wavelength calibration using arc lamp spectra was then checked against sky lines, and in the final pass all spectra were shifted by $\sim$few \AA, so that the [\ion{O}{i}] $\lambda$~6300 sky line was at its rest wavelength. This was done to ensure that all spectra were on a common wavelength scale in the critical region around H$\alpha$ where Gr\#11 and Gr\#16 overlap. 

Low-resolution spectra were obtained with the FLOYDS spectrograph, mounted on the 2-m Faulkes South telescope at Siding Spring Observatory, Australia. These spectra were reduced using the FLOYDS pipeline \footnote{\url{https://github.com/LCOGT/floyds_pipeline}} \citep{Valenti2014}. The automatic reduction pipeline splits the first and second order spectra into red and blue arms and rectifies them using a Legrendre Polynomial. Data is then trimmed, flat-fielded using images taken during the observing block and cleaned of cosmic rays. Red and blue arms are then flux and wavelength calibrated and then merged into a 1D spectrum.

A single spectrum was obtained with the WiFeS IFU spectrograph, mounted on the ANU 2.3m telescope. This spectrum was reduced with the \textsc{PyWiFeS} pipeline \citep{Chil14}. 

\begin{figure*}
\centering
\includegraphics[width = \textwidth]{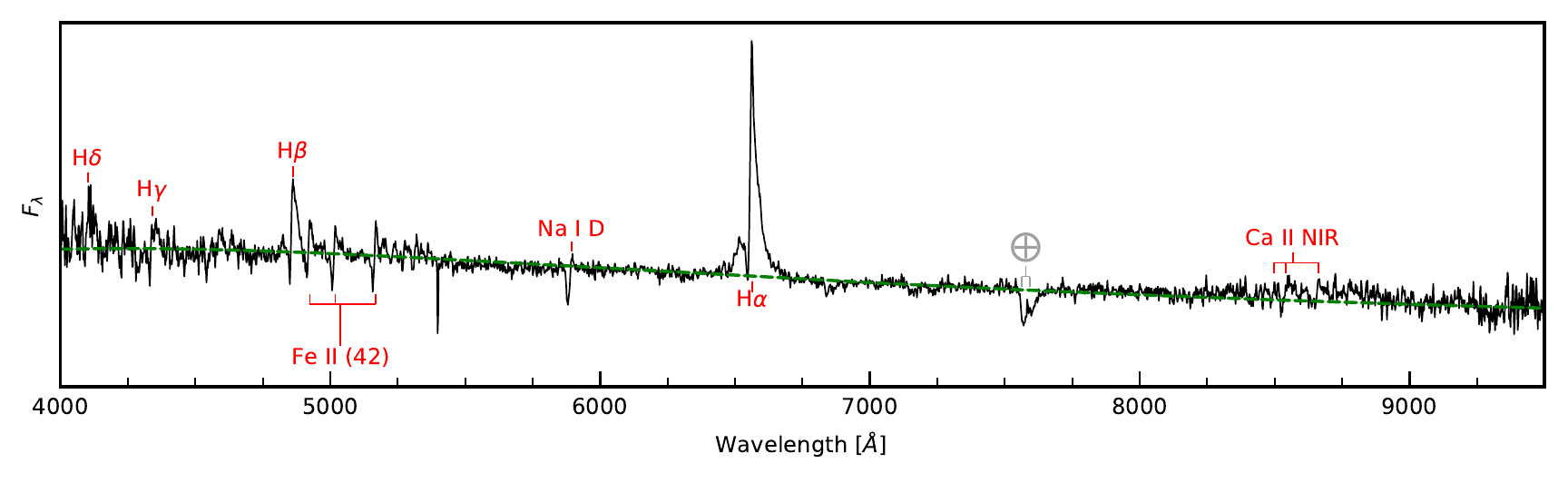}
\caption{Classification spectrum of \jbu obtained with the Du Pont 2.5-m telescope and WFCCD \citep[and reported in][]{Bose17} taken on 2016 December 31 ($-31.4$~d), corrected for reddening. This spectrum coincides with the approximate peak of \textit{Event A}. The green dashed line is the blackbody fit with $T_{BB}$ $\sim$~6750~K. H$\alpha$ and H$\beta$ dominate the spectra and are both well fitted with a P Cygni profile with an additional emission component. We can also distinguish the Na~I~D lines superimposed on \ion{He}{I} $\lambda~5875$ absorption. \ion{Fe}{ii} $\lambda\lambda$ 4924,5018,5169 are present, all with a P Cygni profiles, giving a velocity at maximum absorption of $\sim-700$~\kms. A noise spike at 5397~\AA\ has been removed manually.}
\label{fig:dupont_spectra}
\end{figure*}

All optical spectra are listed in Table \ref{tab:opticalspec} and are shown in Fig.~\ref{fig:all_spectra}. For completeness, we also include the classification spectrum of \jbu in our analysis obtained with the du Pont 2.5-m telescope + WFCCD \citep[and reported in][]{Bose17}, as it is the earliest spectrum available of the transient, see also Fig.~\ref{fig:dupont_spectra}.

\begin{table*}

\caption{Log of optical, UV, and NIR spectra obtained for \jbu. MJD refers to the start of the exposure. Phase is with respect to the time of $V$-band maximum (MJD $57784.4\pm0.5$).}
\setlength{\tabcolsep}{4pt} 
\label{tab:opticalspec}
\begin{threeparttable}
\begin{tabular}{cclcc}
 \hline
Date & MJD & Phase (days) & Instrument & Grism \\
 \hline
2016-12-31 & 57753.0 & $-31.4$ & DuPont+WFCCD & Blue grism \\
2017-01-02 & 57755.4 & $-28.0$ & Magellan+FIRE & LDPrism \\
2017-01-04 & 57757.3 & $-27.1$ & NTT+EFOSC2 & Gr\#13 \\
2017-01-06 & 57759.3 & $-25.1$ & NTT+EFOSC2 & Gr\#11 \\
2017-01-08 & 57761.7 & $-22.7$ & FTS+FLOYDS & red/blue \\
2017-01-15 & 57768.5 & $-15.9$ & FTS+FLOYDS & red/blue \\
2017-01-17 & 57770.2 & $-14.2$ & NTT+EFOSC2 & Gr\#11+Gr\#16 \\
2017-01-18 & 57771.3 & $-13.1$ & NTT+EFOSC2 & Gr\#11+Gr\#16 \\
2017-01-20 & 57773.2 & $-11.2$ & NTT+EFOSC2 &  Gr\#11 \\
2017-01-20 & 57773.1 & $-10.7$ & Gemini S+FLAMINGOS2 &  JH \\
2017-01-22 & 57775.2 & $-9.2$ & {\it Swift} + UVOT & UV Grism \\
2017-01-26 & 57779.3 & $-5.1$ & NTT+EFOSC2 & Gr\#11+Gr\#16 \\
2017-01-27 & 57780.0 & $-4.4$ & ANU 2.3m+WiFeS & red/blue \\
2017-01-27 & 57780.2 & $-4.2$ & NTT+EFOSC2 & Gr\#11+Gr\#16 \\
2017-01-27 & 57780.7 & $-3.7$ & FTS+FLOYDS & red/blue \\
2017-01-28 & 57781.2 & $-3.2$ & NTT+EFOSC2 & Gr\#11+Gr\#16 \\
2017-01-30 & 57783.6 & $-0.8$ & FTS+FLOYDS & red/blue \\
2017-02-02 & 57786.3 & +1.9 &  Gemini S+FLAMINGOS2 &  JH \\
2017-02-02 & 57786.5 & +2.1 & FTS+FLOYDS & red/blue \\
2017-02-04 & 57788.4 & +4.0 & NTT+EFOSC2 & Gr\#11+Gr\#16 \\
2017-02-07 & 57791.2 & +6.8 & NTT+EFOSC2 & Gr\#11+Gr\#16 \\
2017-02-08 & 57792.6 & +8.2 & FTS+FLOYDS & red/blue \\
2017-02-11 & 57795.7 & +11.3$^*$ & FTS+FLOYDS &  red \\
2017-02-14 & 57798.5 & +14.1 & FTS+FLOYDS & red/blue \\
2017-02-17 & 57801.5 & +17.1 & FTS+FLOYDS & red/blue \\
2017-02-19 & 57803.2 & +18.8 & NTT+EFOSC2 & Gr\#11+Gr\#16 \\
2017-02-20 & 57804.6 & +20.2 & FTS+FLOYDS & red/blue \\
2017-02-24 & 57808.6 & +24.2 & FTS+FLOYDS & red/blue \\
2017-02-25 & 57809.1 & +24.7 & NTT+EFOSC2 & Gr\#11+Gr\#16 \\
2017-02-27 & 57811.1 & +26.7 & NTT+EFOSC2 & Gr\#11+Gr\#16 \\
2017-03-06 & 57818.1 & +33.7 & NTT+EFOSC2 & Gr\#11+Gr\#16 \\
2017-03-06 & 57818.5 & +34.1$^*$ & FTS+FLOYDS & red/blue \\
2017-03-11 & 57823.5 & +39.1$^*$ & FTS+FLOYDS &  red \\
2017-03-24 & 57836.0 & +51.6 & NTT+EFOSC2 & Gr\#11+Gr\#16 \\
2017-03-28 & 57840.5 & +56.1 & FTS+FLOYDS &  red \\
2017-04-01 & 57844.5 & +60.1 & FTS+FLOYDS &  red \\
2017-04-14 & 57857.5 & +73.1 & FTS+FLOYDS & red/blue \\
2017-04-22 & 57865.0 & +80.6$^*$ & NTT+EFOSC2 & Gr\#11+Gr\#16 \\
2017-05-01 & 57874.1 & +89.7 & NTT+EFOSC2 & Gr\#11+Gr\#16 \\
2017-06-01 & 57905.1 & +120.7 & NTT+EFOSC2 & Gr\#11+Gr\#16 \\
2017-08-21 & 57986.3 & +201.9 & NTT+EFOSC2 & Gr\#11+Gr\#16 \\
2017-08-22 & 57987.3 & +202.9 & NTT+EFOSC2 &  Gr\#16 \\
2017-09-29 & 58025.3 & +240.9 & NTT+EFOSC2 & Gr\#11+Gr\#16 \\
2017-10-28 & 58054.3 & +269.9 & NTT+EFOSC2 & Gr\#11+Gr\#16 \\
2017-11-26 & 58083.3 & +298.9 & NTT+EFOSC2 & Gr\#11+Gr\#16 \\
2018-01-12 & 58130.2 & +345.8 & NTT+EFOSC2 & Gr\#11+Gr\#16 \\
2018-02-19 & 58168.3 & +383.9 & NTT+EFOSC2 & Gr\#11+Gr\#16 \\
2018-03-26 & 58203.1 & +418.7 & NTT+EFOSC2 & Gr\#11+Gr\#16 \\
\hline
\end{tabular}
\begin{tablenotes}\footnotesize
\item[*] Spectrum not plotted in Fig.~\ref{fig:lc} due to low S/N but still used in analysis when applicable for Fig.~\ref{fig:halpha_param_evo}.
\end{tablenotes}
\end{threeparttable}
\end{table*}

We present a single NIR spectrum taken in the low-dispersion and high-throughput prism mode with FIRE \citep{Simc13} mounted on one of the twin Magellan Telescopes (Fig.~\ref{fig:IR_spectra}). The spectrum was obtained using the ABBA ``nod-along-the-slit'' technique at the parallactic angle. Four sets of ABBA dithers totalling 16 individual frames and 2028.8s of on-target integration time were obtained. Details of the reduction and telluric correction process are outlined by \citet{Hsiao19}. 

In addition, we present two spectra taken with Gemini South + Flamingos2 \citep{Eike04} in long-slit mode. An ABBA dither pattern was used for observations of both \jbu and a telluric standard. These data were reduced using the {\sc gemini.f2} package within {\sc iraf}. A preliminary flux calibration was made using the telluric standard on each night (in both cases a Vega analog was observed), and this was then adjusted slightly to match the $J-H$ colour of \jbu from contemporaneous NIR imaging.

{\it Swift}+UVOT spectra were reduced using the {\sc uvotpy} {\sc python} package \citep{Kuin_2014} and calibrations from \citet{Kuin_2015}.

\section{Photometric evolution}\label{sec:phot_evo}

\subsection{Overall evolution}

We present our complete lightcurve for \jbu in Fig.~\ref{fig:lc}, spanning from $\sim$~10 years before maximum brightness (MJD: 57784.4) to $\sim$~1.5 years after maximum light. \citetalias{Kilpatrick2018} mainly focuses on the time around maximum light up until +118 days. on \jbu. Our photometric coverage is much higher cadence and covers a wider wavelength range.

For the purpose of discussion, we adopt the nomenclature for features seen in the lightcurve of \ip from \citet{Graham2014}; rise, decline, knee, and ankle. We do not designate a ``bump'' phase as while \ip shows a clear bump at $\sim$~20~d, this is not seen in \jbu. The rise begins at $\sim~+22$ days prior to $V$-band maximum. The \textit{decline} phase begins at $V$-band maximum. The \textit{plateau} begins at $\sim~+20$ days, when the decline gradient flattens out initially. The \textit{knee} stage is $\sim~+45$ days past maximum when a sharp drop is seen in the lightcurve, and the \textit{ankle} is the flattening of the lightcurve after $\sim$~65 days before the seasonal gap.

\jbu shows a clear double-peaked lightcurve which has been previously missed in literature. The first fainter peak (at MJD 57751.2, mainly seen in \textit{r}-band) will be referred to as ``\textit{Event A}'', and the subsequent brighter peak is ``\textit{Event B}''. \textit{Event A} is first detected around three months (phase: $-91$~d) before the \textit{Event B} maximum in VLT+FORS2 imaging \citep{atel9938}. Phases presented in this paper for \jbu and other \ip-like transients will always be in reference to \textit{Event B} maximum light (MJD 57784.4). The rise and decline of this first peak is clearly seen in \textit{r}-band (mainly detected from the Prompt telescope array) and sparsely sampled by {\it Gaia} in \textit{G}-band. \textit{Event A} has a rise time to peak of $\sim$~60~days, reaching an apparent magnitude \textit{r} $\sim~18.12$~mag (absolute magnitude $-13.96$~mag). We then see a short decline in \textit{r}-band for $\sim$~2 weeks until \jbu exhibits a second sharp rise seen in all photometric bands, starting on MJD 57764. 

We regard the start of this rise as the beginning of \textit{Event B}. The second event has a faster rise time of $\sim$~19 days, peaking at \textit{r}$\sim$~13.8~mag (absolute magnitude $-18.26$~mag). Our high cadence data shows after $\sim$~20 days past the \textit{Event B} maximum, a flattening is seen in Sloan-\textit{gri} and Cousins \textit{BV} that persists for $\sim$~2 weeks, with a decline rate $\sim0.04~\si{mag~d^{-1}}$. At $\sim$~50 days, a rapid drop is seen at optical wavelengths, with the drop being more pronounced in the redder bands and less in the bluer bands. After the drop there is a second flattening. After two months from the \textit{Event B} peak, the optical bands flatten out with a decay of $\sim0.015$~mag~d$^{-1}$ and remain this way until the seasonal gap at $\sim$~120 days. 

Our dataset includes late time coverage of \jbu not previously covered in the literature. A re-brightening event is seen after $\sim$~120 days and is seen clearly in \textit{BVGgr}-bands. We miss the initial rebrightening event in our ground-based data, so it is unclear if this is a plateau lasting across the seasonal gap or a re-brightening event. However, evidence for a rebrightening in the lightcurve is seen in {\it Gaia}-\textit{G} (See Fig.~\ref{fig:lc}). We can deduce that this event occurred between +160 and +195 days from our {\it Gaia}-\textit{G} data, where we have $G = 18.69$~mag at +160 days, but an increase to 18.12~mag one month later. An additional bump is seen in {\it Gaia}-\textit{G} at +345 days. We observe $G = 18.95$~mag at +316~d and $G=18.88$~mag at +342~d before \jbu fades to $G=19.72$~mag a month later. 

Late time bumps and undulations in the lightcurves of SNe are commonly associated with late time CSM interaction, when SN ejecta collide with dense stratified and/or clumpy CSM far away from the progenitor, providing a source of late time energy \citep{Fox2013,Martin_2014,Arcavi2017,Nyholm2017,Andrews2018,Moriya2020}.

\subsection{Color Evolution}\label{sec:color_comp}
There exists a growing sample of \ip-like transients which evolve almost identically in terms of their photometry and spectroscopy, in the years prior to, and during their main luminous events.
The color evolution of \jbu is discussed by \citetalias{Kilpatrick2018}. However, we include color information prior to \textit{Event B} maximum. Additionally we show late-time color evolution of K18.
In addition to \jbu, we focus on a small sample of objects that show common similarities to \jbu. For the purpose of a qualitative study, we will compare \jbu with \ip \citep{Fraser13, Graham2014}, \bh \citep{Elisa-Rosa2016,Thone2017}, \lsq \citep{Tartaglia2016}, \gc \citep{Reguitti2018} and \bdu \citep{Pastorello_2017}. We will refer to these transients (including \jbu) as \ip-like transients.
We also include \al \citep{Benetti_2016} in our \ip-like sample. Although no pre-explosion variability or an \textit{Event A/B} lightcurve was detected, \al shows a similar bumpy decay from maximum and a similar spectral evolution as well as showing no sign of explosively nucleosynthesized material; e.g. $[\ion{O}{I}]~\lambda\lambda~6300,6364$ even after 15 years. A modest ejecta mass and restrictive constraint on the ejected $^{56}$Ni mass are similar to what is found for \jbu and other \ip-like transients, see \paperII. \citet{Benetti_2016} suggest that this is consistent with a fall-back supernova in a highly structured environment, and we discuss this possibility for \jbu in \paperII.
We will also discuss \cnf \citep{Pastorello2019}; a previously classified Type IIn SN \citep{ATel11726}. Although \citet{Pastorello2019} argues that \cnf displays many of the characteristics of \ip, it does not show the degree of asymmetry in H$\alpha$ when compared to \jbu but does show pre-explosion variability and general spectral evolution similar to \ip-like transients.
\begin{figure*}
\centering
\includegraphics[width=\textwidth]{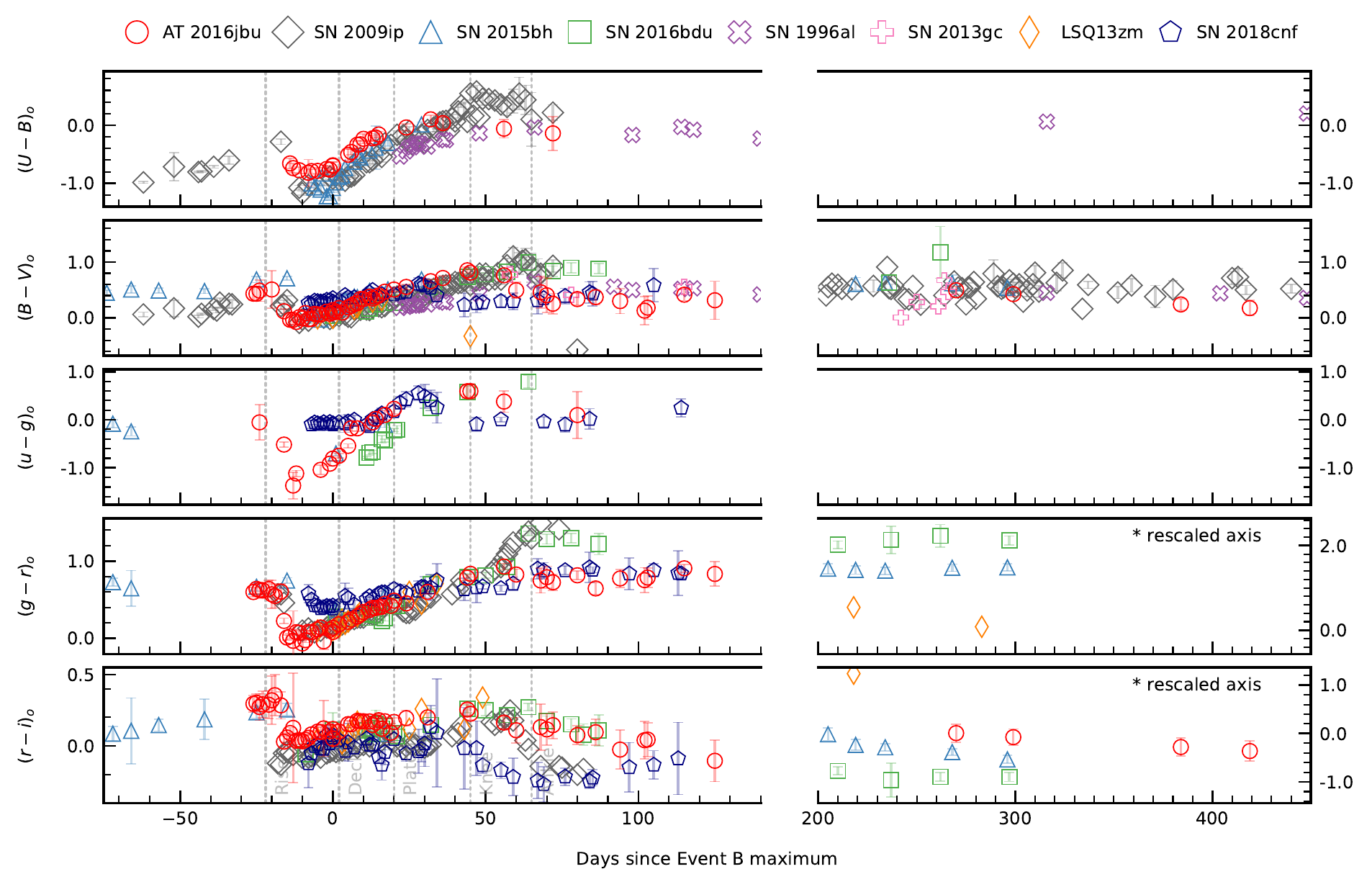}
\caption{Intrinsic color evolution of \jbu and \ip-like transients. All transients have been corrected for extinction using the values from Table.~\ref{tab:objects}. X-axis gives days from \textit{Event B} maximum light. We include a broken X-axis to exclude the seasonal gap for \jbu. Data shown for \jbu has been regrouped into 1 day bins and weighted averaged. Error bars are shown for all objects, and we do not plot any point with an uncertainty greater than 0.5 mag. The different stages of evolution of \jbu are marked with grey dashed vertical bands.}
\label{fig:compare_color}
\end{figure*}
Fig.~\ref{fig:compare_color} shows that all these transients show a relatively slow color evolution, typically seen in Type IIn SNe \citep{Taddia_2013,Nyholm_2020}. Where color information is available, \ip-like transients initially appear red $\sim$~1 month before maximum light, becoming bluer as they rise to maximum light. This is best seen in $(B-V)_0$ for \jbu, \bh and \ip. These three transients span colors from $(B-V)_0 \sim$0.5 at $\sim-20$~d to $\sim$0.0 at $\sim-10$~d. In general, after the peak of \textit{Event B} the transients begin to cool and again evolve towards the red.
For the first $\sim$~20~days after \textit{Event B}, \jbu follows the trend of other transients, which is seen clearly in $(U-B)_0$, $(B-V)_0$, $(g-r)_0$, and $(r-i)_0$. At $\sim$~20~d \jbu flattens in $(U-B)_0$ and $(r-i)_0$, similar to \al and \cnf, whereas \ip flattens at $\sim$~40~d in $(U-B)_0$. This phase corresponds with the \textit{plateau} stage in \jbu. This feature is also seen in $(r-i)_0$ and $(u-g)_0$, where \jbu plateaus at $\sim$~20~d and then slowly evolves to the blue.
This behaviour is also seen in $(B-V)_0$ and $(g-r)_0$, where a color change is observed at $\sim$~50~d, followed by \jbu remaining at approximately constant color until the seasonal gap at $\sim$~120~d.
\cnf follows the trend of \jbu quite closely in $(B-V)_0$ but this abrupt transition to the blue is seen at $\sim$~30~d in \cnf, and $\sim$~60~d in \jbu. \jbu and \cnf are distinct in their $(g-r)_0$ evolution, as they match \ip and \bdu closely until $\sim$~50~d, after which \jbu remains at an approximately constant color, while \ip and \bdu make an abrupt shift to the red.
Filters that cover H$\alpha$ (viz. \textit{r,V}) show an abrupt color change at $\sim$~60~d in \jbu (i.e $(B-V)_0$, $(g-r)_0$, and $(r-i)_0$), whereas those that do not cover H$\alpha$ show a similar feature at $\sim$~30~d i.e. $(U-B)_0$ and $(u-g)_0$. As noted by \citetalias{Kilpatrick2018}, at this time we see an increase in the relative strength of the H$\alpha$ blue shoulder emission component (see Sect.~\ref{sec:balmer_evo}). $(B-V)_0$, $(g-r)_0$, and $(r-i)_0$ do not show this trend but rather a transition to the blue at $\sim$~60~d. At late times, $>120$~days, \jbu remains relatively blue and follows the trends of other \ip-like transients, especially in $(B-V)_0$.
%

\subsection{Ground Based Pre-Explosion Detections}

A trait of \ip-like transients is erratic photometric variability\footnote{referred to as ``\textit{flickering}'' in \citealp{Kilpatrick2018}.} in the period leading up to \textit{Event A} and \textit{Event B}.

The lower panel of Fig.~\ref{fig:lc} shows all pre-\textit{Event A/B } observations for \jbu from ground based instruments. The majority of these observations are from the PROMPT telescope array, and have been host subtracted using late time $r$-band templates from EFOSC2. Unfortunately, these images are relatively shallow. In addition, we recovered several images from the LCO network which were obtained for the follow-up campaign of SN~2015F \citep{Cartier17}. These images have been host subtracted using templates from LCO taken in 2019. We also present several images taken VLT+OMEGAcam which are deeper than our templates and are hence not host subtracted. For completeness we also plot detections of the progenitor of \jbu from {\it HST} in Fig.~\ref{fig:lc}, which we discuss in \paperII.

\begin{figure}
\centering
\includegraphics[width=\columnwidth]{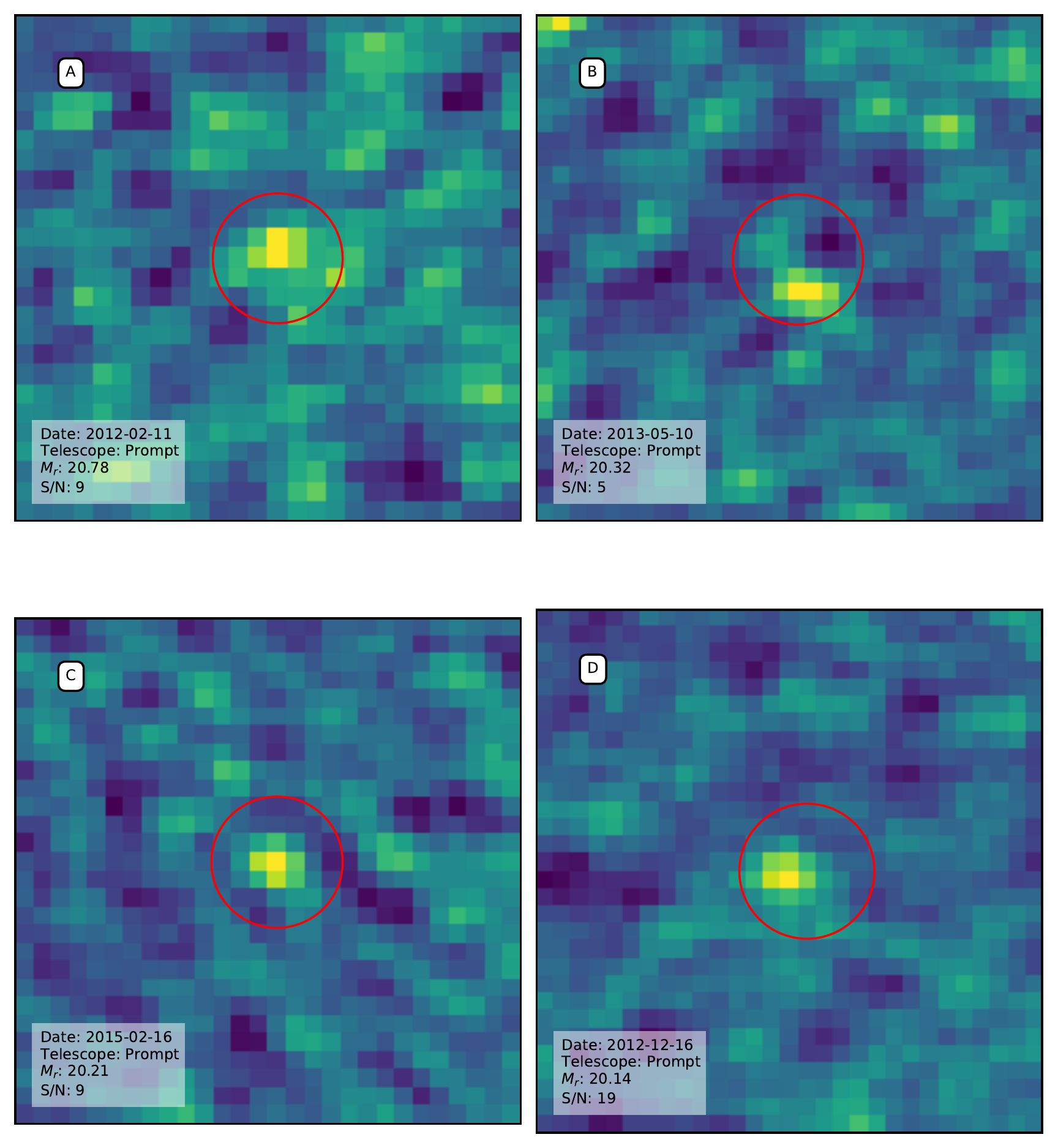}
\caption{Sample of pre-explosion detections from PROMPT at the progenitor location. Center of cutout corresponds to \jbu progenitor location. Red circle signifies aperture with radius $1.3 \times$ FWHM placed in the center of the cutout. As mentioned in Sect.~\ref{sec:image_reduction}, these unfiltered images have been host subtracted using \textit{r} band templates. Template subtractions performed using \apt and {\sc HOTPANTS} \citep{hotpants}, see Sect.~\ref{sec:autophot}.}
\label{fig:prompt_detections}
\end{figure}

If \jbu underwent a similar series of outbursts prior to \textit{Event A/B} as seen in other \ip-like transients, then we would expect to only detect the brightest of these. \ip experiences variability at least three years prior to its main events. 

For \jbu, several significant detections are found with \textit{r}$\sim$~20~mag in the years prior to \textit{Event A/B}. For our adopted distance modulus and extinction parameters, these detections correspond to an absolute magnitude of $M_r\sim-11.8$~mag. Similar magnitudes were seen in \ip and \bh, see Fig.~\ref{fig:compare_lc}. \ip was observed with eruptions exceeding $R\sim-11.8$~mag, with even brighter detections for \bh.

Both \ip and \bh show a large increase in luminosity $\sim$~450~d days prior to their \textit{Event A/B}. The \jbu progenitor is seen in {\it HST} images around $-400$~d showing clear variations. A single DECam image in \textit{r} band gives a detection at \textit{r}$\sim22.28\pm0.26$~mag at $-352$~d which roughly agrees with our \textit{F350LP} lightcurve at this time (if we presume H$\alpha$ is the dominant contributor to the flux). We present and further discuss {\it HST} detections in \paperII.

We note that we detect a point source at the site of \jbu in several PROMPT images but not in any of the LCO, WFI, NTT+EFOSC2/SOFI, OmegaCAM or VISTA+VIRCAM pre-explosion images. However, a clear detection is made with CTIO+DECAM that is compatible with our {\it HST} observations (see \paperII for more discussion of this).

In Fig.~\ref{fig:prompt_detections} we show a selection of cutouts from our host subtracted PROMPT images, showing the region around \jbu. While some of the detections that \apt recovers are marginal, others are quite clearly detected, and so we are confident that the pre-discovery variability is real. If these are indeed genuine detections, then \jbu is possibly undergoing rapid variability similar to \ip and \bh in the years leading up to their \textit{Event A}. The high cadence of our PROMPT imaging and the inclusion of H$\alpha$ in the \textit{Lum} filter plausibly explain why we have not detected the progenitor in outburst in data from any other instrument.

\jbu could be undergoing a slow rise up until the beginning of \textit{Event A} similar to UGC 2773-OT \citep{Smith_2015} (Intriguingly this is also seen in Luminous Red Novae, \citealp{pastorello2020,Williams2015} - we return to this in \paperII). Fitting a linear rise to the PROMPT pre-explosion detections (i.e. excluding the {\it HST} and DECam detections) gives a slope of $-5.4\pm1\times 10^{-4}$~mag~d$^{-1}$ and intercept of $19.07\pm0.19$~mag. If we extrapolate this line fit to $-$60~d (roughly the beginning of \textit{r}-band coverage for \textit{Event A}) we find a value of $r_{extrapolate}\sim19.11$~mag which is very similar to the detected magnitude at $-59$~d of $r\sim19.09$~mag. However, this is speculative, and accounting for the sporadic detections in the preceding years, and the non-detections in deeper images e.g. from LCO see lower panel of Fig.~\ref{fig:lc}, it is more likely that \jbu is undergoing rapid variability (similar to \ip) which is serendipitously detected in our PROMPT images due to their high cadence.

\subsection{UV Observations}\label{sec:UV_obs}

\begin{figure}
\centering
\includegraphics[width=\columnwidth]{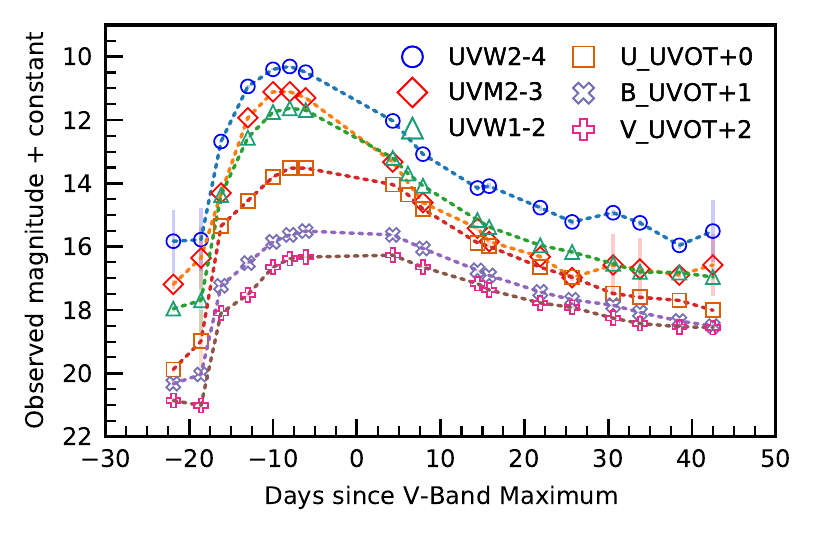}
\caption{{\it Swift} + UVOT lightcurve for \jbu. All photometry is host subtracted. Offsets are given in the legend and uncertainties are included for all points.}
\label{fig:swift_lc}
\end{figure}

Fig.~\ref{fig:swift_lc} shows {\it Swift}+UVOT observations around maximum light. All bands show a sharp increase at $\sim-18$~d, consistent with our optical lightcurve. The {\it Swift}+UVOT can constrain the initial \textit{Event B} rise to some time between $\sim-18.6$~d and $\sim-16.2$~d.

The decline of the UV lightcurve is smooth and does not show any obvious features up to +45~d. {\it UVW2} shows a possible bump beginning at $\sim$~24~d that spans a few days. This bump is also evident in {\it UVM2} at the same time. This bump is consistent with the emergence of a blue shoulder emission in H$\alpha$ (See Sect. \ref{sec:balmer_evo}) and it is possible that we are seeing an interaction site between ejecta and CSM at this time. 

\subsection{X-ray Observations}\label{sec:xray_obs}

No clear X-ray source was found consistent with the location of \jbu in the XMM data taken at -5~d. Using the {\sc sosta} tool on the data from the PN camera we obtain a $3\sigma$ upper limit of $<3.2\times10^{-3}$ counts~s$^{-1}$ for \jbu; while the summed MOS1+MOS2 data gives a limit of $<2.1\times10^{-3}$ counts~s$^{-1}$. Assuming a photon index of 2, the upper limit to the observed flux in the 0.2--10 keV energy range is $1.2 \times 10^{-14}$~erg cm$^{-2}$ s$^{-1}$.

For comparison, \ip was detected in X-rays in the 0.3--10 keV energy band with a flux of $(1.9 \pm 0.2) \times 10^{-14}$~erg cm$^{-2}$ s$^{-1}$, as well as having an upper limit on its hard X-ray flux around optical maximum \citep{Margutti2014}. 

X-ray observations can tell us about the ejecta-CSM interaction as well as the medium into which they are expanding into \citep{Dwarkadas2012}. The non-detection for \jbu provides little information on the nature of \textit{Event A/B}. Making a qualitative comparison to \ip we note that \jbu is not as X-ray bright, and this may reflect different explosion energies, CSM environments or line-of-sight effects.

\subsection{MIR evolution}\label{sec:spitzer}

We measure fluxes for \jbu in {\it Spitzer} IRAC $Ch1$ = 0.123$\pm$0.003~mJy and $Ch2$ = 0.136$\pm$0.003~mJy, which are roughly consistent with those found by \citetalias{Kilpatrick2018}. This corresponding to magnitudes of 16.00 and 15.25 for $Ch1$ and $Ch2$ respectively. Neither this work nor \citetalias{Kilpatrick2018} finds evidence for emission from cool dust in $Ch3$ and $Ch4$ at the progenitor site of \jbu.

We further discuss the evidence for a dust enshrouded progenitor in \paperII but here we briefly report the findings from \citetalias{Kilpatrick2018}. Coupled with pre-explosion {\it HST} observations, \citetalias{Kilpatrick2018} finds that the progenitor of \jbu is consistent with the progenitor system having a significant IR excess from a relatively compact, dusty shell. The dust mass in the immediate environment of the progenitor system
is small (a few $\times10^{-6}$ \msun). However, the different epochs of the {\it HST} (taken in 2016) and {\it Spitzer} (taken in 2003) data suggest they may be at different phases of evolution. Fig.~\ref{fig:lc} shows that the site of \jbu underwent multiple outbursts between 2006 and 2013, and, as mentioned by \citetalias{Kilpatrick2018}, fitting a single SED to the {\it HST} and {\it Spitzer} datasets may be somewhat misleading.

\section{Spectroscopy}\label{sec:spec}

\begin{figure*}
\includegraphics[width = \textwidth]{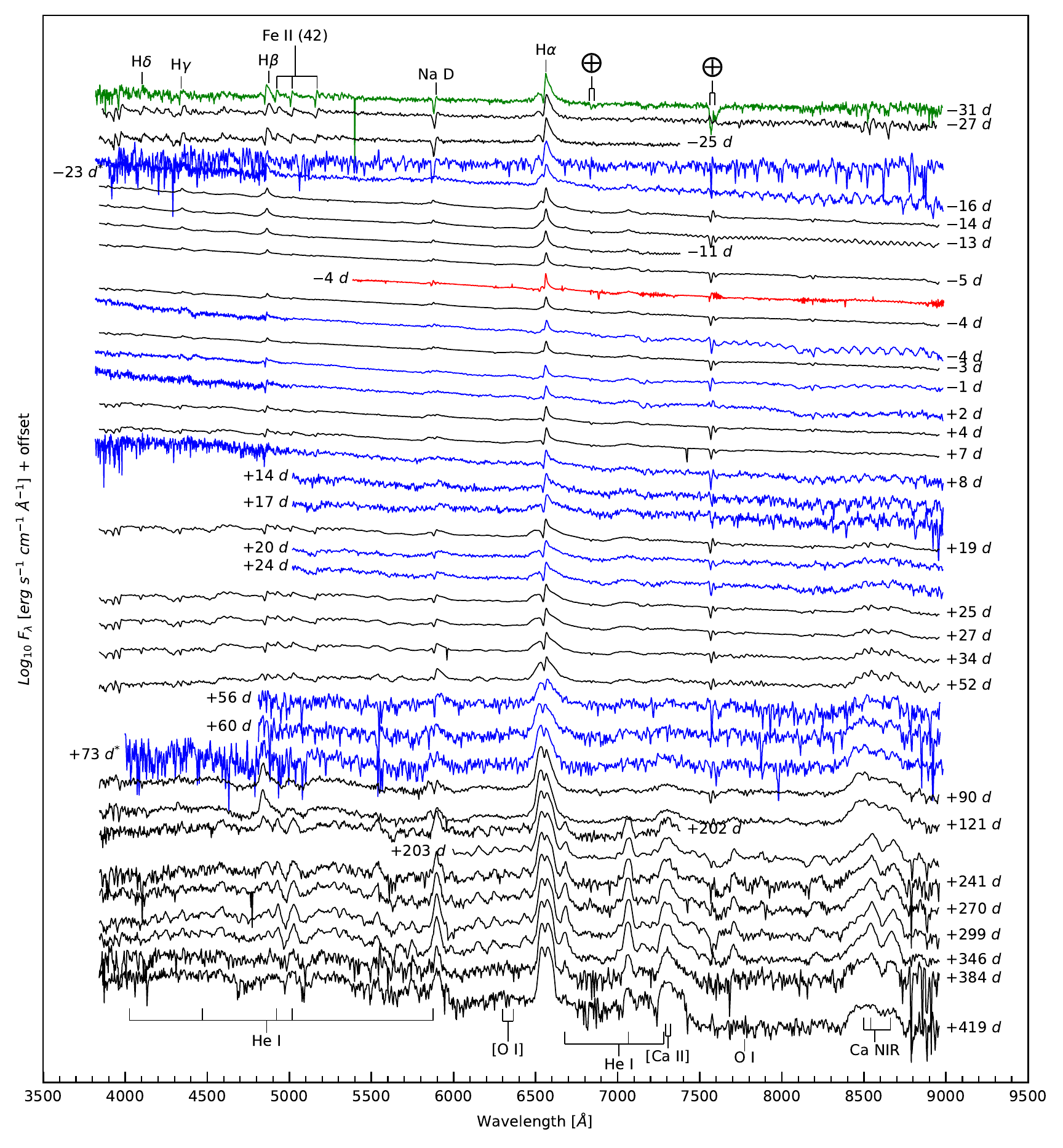}
\caption{Spectral evolution of \jbu. Wavelength given in rest frame. Flux given in log scale. Prominent spectral lines and strong absorption bands are labelled. Colors instruments used (see Table \ref{tab:opticalspec}); black: NTT+EFOSC2, blue: FTS+FLOYDS, red: WiFeS, green: DuPont. Spectra marked with an asterisk have been smoothed using a Gaussian filter of FWHM 1~\AA. }
\label{fig:all_spectra}
\end{figure*}

We present our high cadence spectral coverage of \jbu in Fig.~\ref{fig:all_spectra}. Our spectra begin at $-31$ days and show an initial appearance similar to a Type IIn SN, i.e. narrow emission features seen in H and a blue continuum. Our first spectra coincide with the approximate peak of \textit{Event A}. After around a week, additional absorption and emission features emerge in the Balmer series, which we illustrate in Fig.~\ref{fig:halpha_spec_evo} and plot the evolution of in Fig.~\ref{fig:halpha_param_evo}. The spectrum does not vary significantly over the first month of evolution aside from the continuum becoming progressively bluer with time. H$\alpha$ shows a P Cygni profile with an emission component with FWHM $\sim$~1000~\kms and a blue shifted absorption component with a minimum at $\sim-600$~\kms. The narrow emission lines likely arise from an unshocked CSM environment around the progenitor. Over time \jbu develops a multi-component emission profile seen clearly in H$\alpha$ that persists until late times. We do not find any clear signs of explosively nucleosynthesised material at late times, and indeed the spectral evolution appears to be dominated by CSM interaction at all times. We discuss the evolution of the Balmer series in Sect.~\ref{sec:balmer_evo}. In Sect.~\ref{sec:calcium_evo} we discuss the evolution of $\ion{Ca}{II}$ features and model late time emission profiles. Sect.~\ref{sec:fe} discusses the evolution of several isolated, strong iron lines. Sect.~\ref{sec:he} discusses the evolution of $\ion{He}{I}$ emission and makes qualitative comparisons between $\ion{He}{I}$ features and the optical lightcurve. We present UV and NIR spectra in Sect.~\ref{sec:UVspectra} and Sect.~\ref{sec:NIRspectra} respectively. 

\subsection{Balmer Line Evolution}\label{sec:balmer_evo}

\begin{figure}
\centering
\includegraphics[width=\columnwidth]{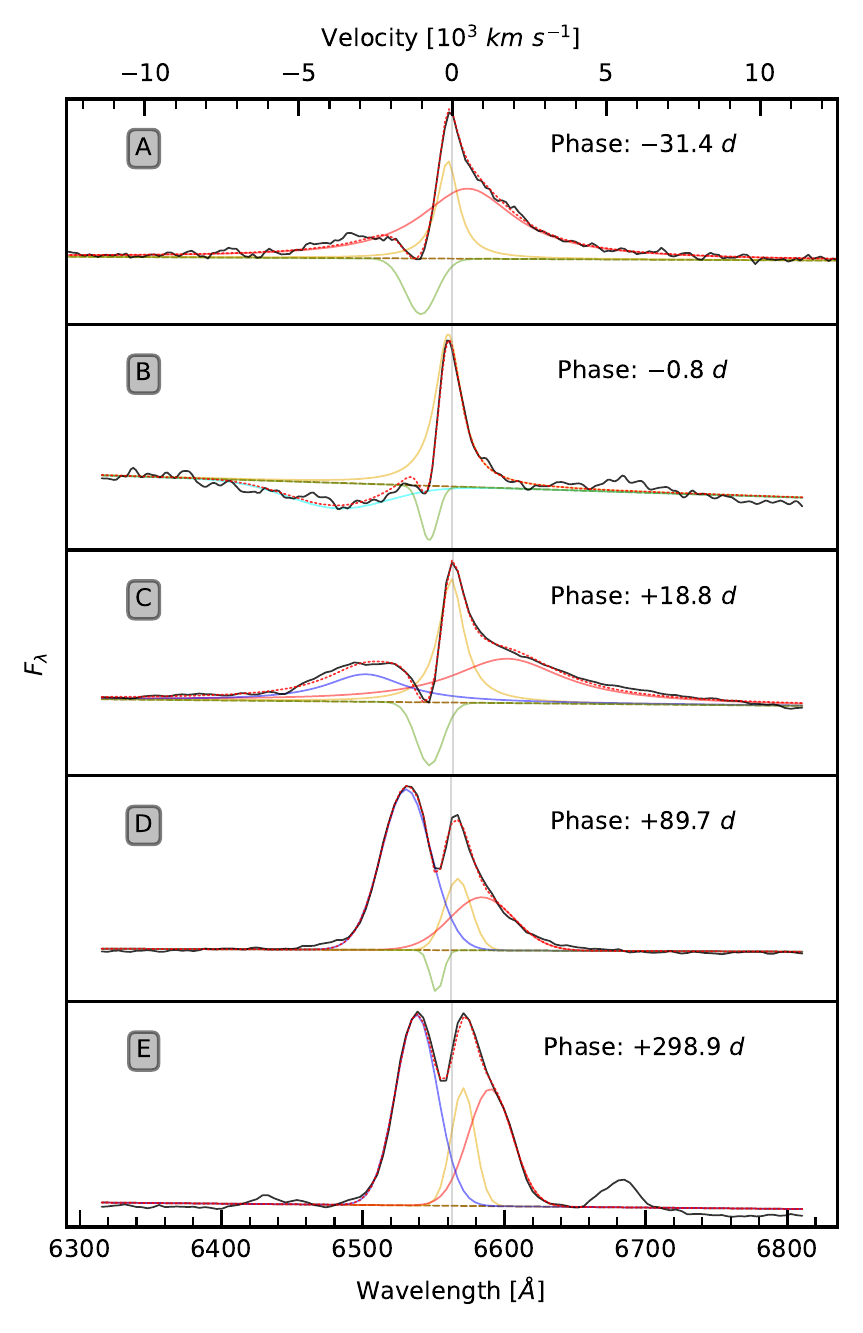}
\caption{Multi-component evolution of H$\alpha$ over a period of $\sim$~1~year. We use Lorentzian emission and Gaussian absorption profiles at early times (phase $<~+120$~d), and Gaussian emission and absorption thereafter. Epochs are given in each panel, lines are coloured such that yellow = core emission, red = redshifted emission, green = P-Cygni absorption, cyan = high velocity absorption and blue is blueshifted emission. In panel A an additional emission component could be included to account for the blue excess shown, although this can simply be extended electron scattering wings.}
\label{fig:halpha_spec_evo}
\end{figure}

The most prominent spectral features are the Balmer lines, which show dramatic evolution over time. In particular the H$\alpha$ profile, which shows a complex, multi-component evolution, provides insight to the CSM environment, mass-loss history and explosion sequence. Although \ip never displayed obvious multi-component emission features, a red-shoulder emission is seen at late times \citep{Fraser13}. We present the evolution of H$\alpha$ for \jbu at several epochs showing the major changes in Fig.~\ref{fig:halpha_spec_evo}.

\citetalias{Kilpatrick2018} discuss the evolution of the H$\alpha$ in detail out to +118~days. With our high cadence spectral evolution we preform a similar multi-component analysis while focusing on individual feature evolution.

\begin{figure}
\centering
\includegraphics[width = \columnwidth]{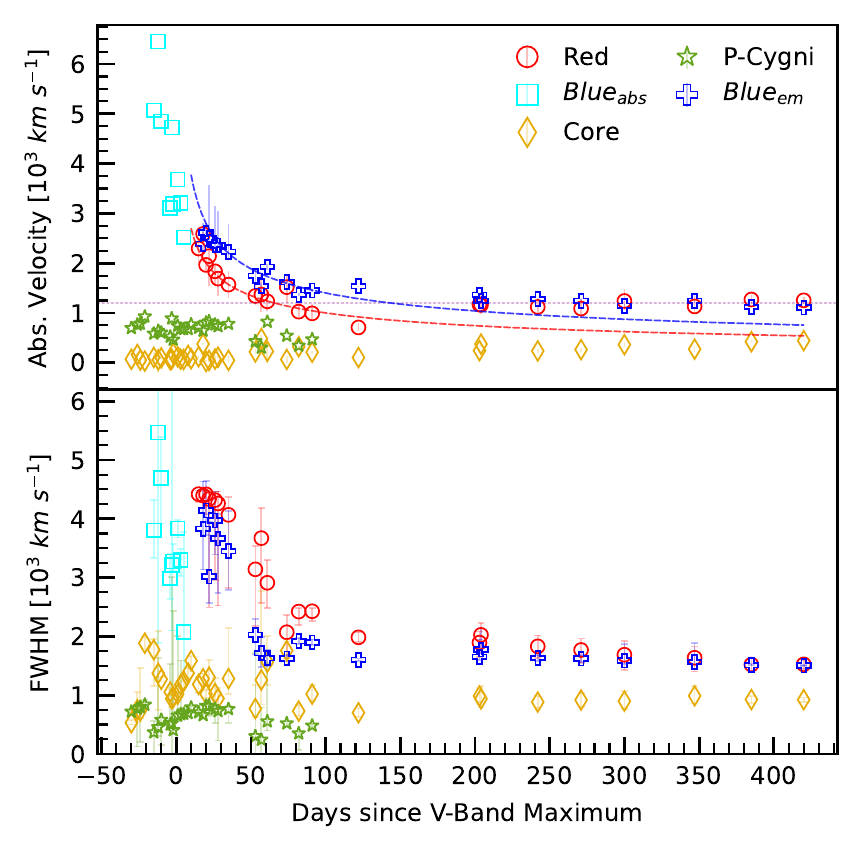}
\caption{Evolution of fitted parameters for H$\alpha$. The upper panel shows the absolute velocity evolution of each feature. We fit a power decay law with index 0.4~dex to the blue emission from when it first appears ($\sim$~+18d) until the seasonal gap ($\sim$~+125~d) indicated by the blue dashed line. The is also fitted for the red shoulder emission (with a different normalisation constant) as the red dashed line. We include a purple dotted line at 1200~\kms that matches the late time red and blue emission components. The lower panel shows the FWHM evolution of each of the components. We do not plot the redshifted broad emission fitted during the first three epochs in either panel.}
\label{fig:halpha_param_evo}
\end{figure}

Similar to \citetalias{Kilpatrick2018}, we conducted spectral decomposition to understand line shape and the ejecta-CSM interaction. We used a Markov Chain Monte Carlo (MCMC) approach for fitting a multi-component spectral profile \citep{lmfit2014} using a custom {\sc python3} script. When fitting, absorption components are constrained to be blueward of the rest wavelength of each line to reflect a P Cygni absorption. All lines are fitted over a small wavelength window and we include a pseudo-continuum during our fitting, which is allowed to vary. Fitting the H$\alpha$ evolution is performed on each spectrum consecutively, using the fitted parameters from the previous model as the starting guess for the next. This is reset after the observing gap at +202~days. Fig.~\ref{fig:halpha_spec_evo} presents fitted models to the H$\alpha$ profile at epochs where significant change are seen. The FWHM and peak wavelength for H$\alpha$ are illustrated in Fig.~\ref{fig:halpha_param_evo}.

{\it Days $-31$ to $-25$}: Similar to \citetalias{Kilpatrick2018}, our first spectrum coincides with the approximate peak of \textit{Event A} (Fig.~\ref{fig:lc}). H$\alpha$ can be modelled by a P Cygni profile with an absorption minimum at $\sim-700$~\kms superimposed on a broad component at $\sim~+700$~\kms with a FWHM of $\sim~2600$~\kms. This can be interpreted as a narrow P Cygni with extended, electron-scattering wings, as often seen in Type IIn SN spectra (see review by \citealp{Filippenko1997}). 

{\it Days $-14$ to $+4$}: We see a gradual decay in amplitude of the core broad emission until we find a best fit by a single intermediate width Lorentzian profile (FWHM $\sim$~1000~\kms) and P Cygni absorption. Our Lorentzian profile has broad wings, possibly due to electron scattering along the line-of-sight \citep{Chugai_2001}. For further discussion, see  \citetalias{Kilpatrick2018}.

At $-14$ days, a blue broad absorption component clearly emerges at $\sim$-5000~\kms with an initial FWHM of $\sim$~3800~\kms, with the fastest material is moving at $\sim$~10,000~\kms. This feature was note seen in \citetalias{Kilpatrick2018} due to a lack of observations at this phase. The trough of this absorption features slows to $\sim$-3200~\kms at $+3$~d. Panel B in Fig.~\ref{fig:halpha_spec_evo}, shows H$\alpha$ at $-1$ days with a strong Lorentzian emission with the now obvious blue absorption. This feature indicates that there is fast moving material that was not seen in the initial spectra. Assuming free expansion, we set an upper limit on the distance travelled by this material to $\sim$\SI{2.5e15}{cm}. 

A similar feature was also seen in \ip, \citep[e.g. Fig. 2 of][]{Fraser13} around the \textit{Event B} maximum. A persistent second absorption feature was also seen in \bh \citep{Elisa-Rosa2016} which remained in absorption until several weeks after the \textit{Event B} maximum, when it was replaced by an emission feature at approximately the same velocity. 

{\it Days $+7$ to $+34$}: A persistent P Cygni profile is still seen but a dramatic change is seen in the overall H$\alpha$ profile, now being dominated by a red-shifted broad Gaussian feature centered at $\sim$+2200~\kms and FWHM $\sim$~4000~\kms. The blue absorption component has now vanished and been replaced with an emission profile with a slightly lower velocity, $-2400$~\kms at +18~d, seen in panel C of Fig.~\ref{fig:halpha_spec_evo}. Over the following month, this line moves towards slower velocities with a decreasing FWHM. The blue shoulder emission is clearly seen at $\sim$+18~d and remains roughly constant in amplitude (with respect to the core component) until $\sim$+34~d. At $+34$~d this line now has a FWHM $\sim$~2700~\kms. By $+52$~d this blue emission line has grown considerably in amplitude with respect to the core component. During this period the relative strength of the red and blue component begins to change, indicating on-going interaction and/or changing opacities. We note that prior to $+52$~d, this H$\alpha$ profile may be fitted with a single, broad emission component with a P Cygni profile. However, during our fitting a significant blue excess was always present during $+7$~d to $+34$~d. Allowing for both a blue and red emission component during these times, allows each consistent component to evolve smoothly into later spectra, as is seen in Fig~\ref{fig:halpha_spec_evo} and Fig.~\ref{fig:halpha_param_evo}.

\begin{figure}
\centering
\includegraphics[width=\columnwidth]{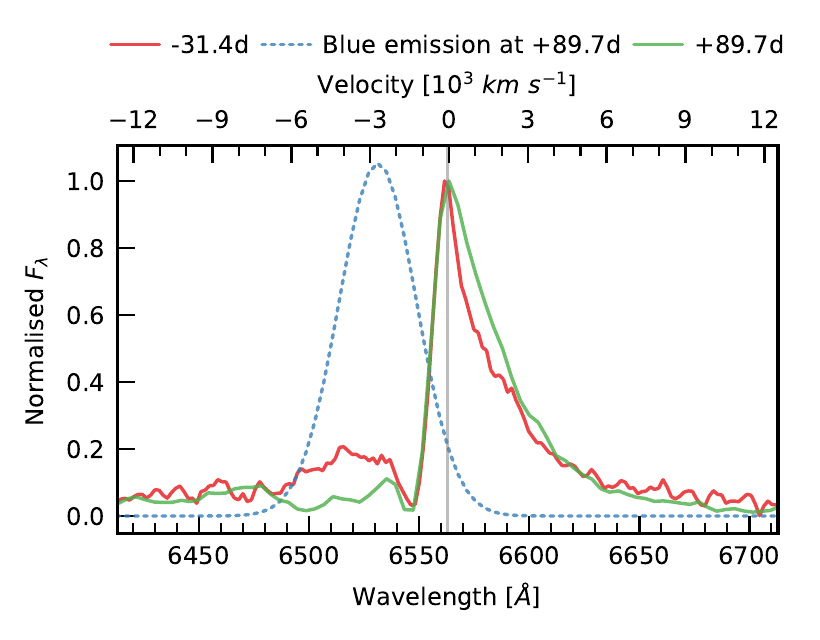}
\caption{H$\alpha$ profile at $-31$~d (red) and $+90$~d (green) for \jbu. The $+90$~d profile has had a strong blue emission profile (given by dotted blue line) subtracted and we plot the residual in green. Each spectra is normalised at 6563~\AA. The profile at $+90$~d has been blue-shifted by 4\AA\ ($\sim-180$~\kms) to match the peak at the H$\alpha$ rest wavelength (6563~\AA) of the profile at $-31$~d.} 
\label{fig:compare_halpha_-30_91}
\end{figure}

{\it Days $+52$ to $+120$}: As mentioned in by \citetalias{Kilpatrick2018}, H$\alpha$ shows an almost symmetric double-peaked emission profile. The earliest profile of H$\alpha$ at $-31$~d is reminiscent of some stages during an eruptive outburst from a massive star \citep[for example Var C;][]{Humphreys_2014}. We plot the profile of the $+90$~d profile in Fig.~\ref{fig:compare_halpha_-30_91} with a blue-shifted Lorentzian profile removed. The profiles are very similar in overall shape with a slightly broader red-core component in the $+90$~d spectrum. A possible interpretation is the P Cygni-like profile seen in our $-31$~d spectra is associated with the events during/causing \textit{Event A} (for example a stellar merger or eruptive outburst) and the blue side emission is associated with events during/causing \textit{Event B} (for example a core-collapse or CSM interaction). 

{\it Days $+203$ to $+420$}: We present late-time spectra of \jbu not previously covered in the literature. The red and blue components of the H$\alpha$ profile now have similar FWHM of $\sim$~2100~\kms and $\sim$~1600~\kms respectively. The overall H$\alpha$ profile has retained its symmetric appearance (panel D of Fig.~\ref{fig:halpha_spec_evo}. After this time we no longer fit a P Cygni absorption profile, and our spectra can be fitted well using three emission components. We justify this as any opaque material may have become optically thin after $\sim$~7~months and the photospheric phase has ended. 

Little evolution in H$\alpha$ is seen for the remainder of our observations. The three emission profiles remain at their respective wavelengths and the approximate same width. The overall evolution of H$\alpha$ suggests that \jbu underwent a large mass loss event (whether that be a SN or extreme mass loss episode) in a highly aspherical environment.Interaction with dense CSM forming a multi-component H$\alpha$ profile as well as a bumpy lightcurve.

\subsection{Calcium Evolution}\label{sec:calcium_evo}

\begin{figure}
\centering
\includegraphics[width=\columnwidth]{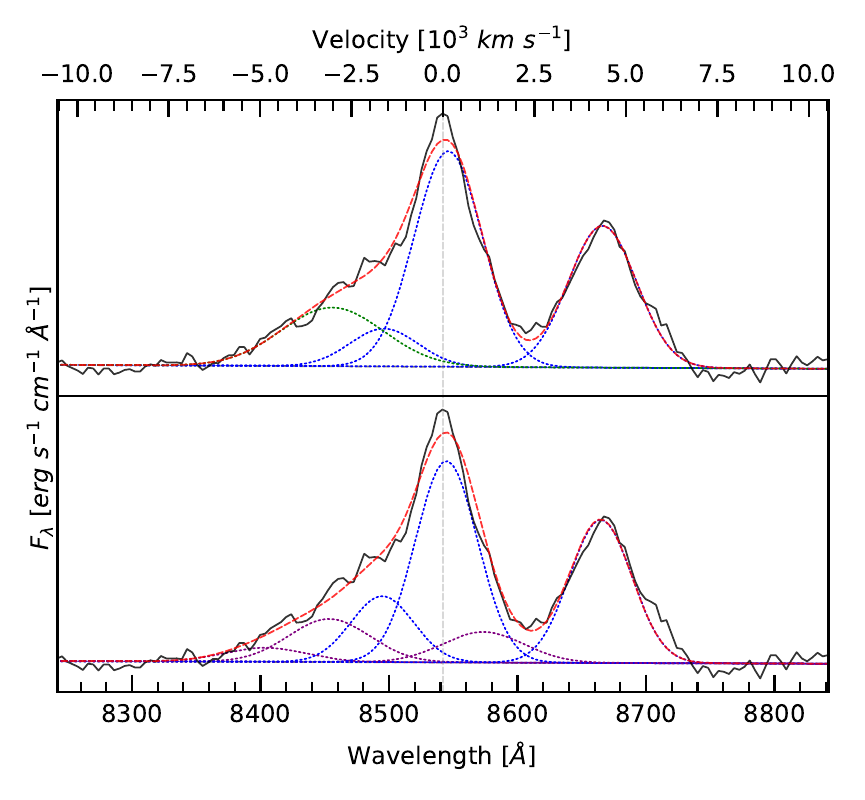}
\caption{Calcium NIR triplet fit for +345~d. The individual components of the primary \ion{Ca}{II} NIR triplet is given by the blue dashed lines in both plots. The upper panel shows the emission profile with the inclusion of $\ion{O}{I}~\lambda8466$ (in green). The lower panel shows the model fit in blue (Region A) with the second region of \ion{Ca}{II} NIR triplet emission shown in purple (Region B). Both $\ion{O}{I}~\lambda8466$ and a second region of Ca emission give a similarly acceptable fit to the data.}
\label{fig:ca_fitting_comparision}
\end{figure}

Sect.~\ref{sec:balmer_evo} indicates that \jbu has a highly non-spherical environment. We investigate similar trends in other emission profiles. \citetalias{Kilpatrick2018} suggest that the [\ion{Ca}{II}] and \ion{Ca}{II} NIR triplet may be coming from separated regions. Motivated by this, we explore the \ion{Ca}{II} NIR triplet $\lambda\lambda\lambda~$ 8498, 8542, 8662 using the same method in Sect.~\ref{sec:balmer_evo}. The \ion{Ca}{II} NIR triplet appears in emission at approximately the same time as blue-shifted emission in H$\alpha$ ($\sim +18$~d) and at early times shows P Cygni absorption minima at velocities similar to H$\alpha$. For profile fitting, the wavelength separation between the three components of the NIR triplet was held fixed, while the three components were also constrained to have the same FHWM. Amplitude ratios between the three lines were constrained to physically plausible values between the optically thin and optically thick regimes \citep{Herbig1980}. 

The early evolution of  the \ion{Ca}{II} NIR triplet is detailed in \citetalias{Kilpatrick2018}.We explore two scenarios for the \ion{Ca}{II} NIR triplet evolution after +200~d. In the first, we assume that the \ion{Ca}{II} emission comes from the same regions as H$\alpha$ (as suggested in Sect.~\ref{sec:balmer_evo}) i.e two spatially separated emitting regions. We allow the first region to be fitted with the above restrictions (fixed line separation, single common FWHM), we refer to this as Region A. A second, kinematically distinct, multiplet is added (we refer to this as Region B) and simultaneously fitted with additional constraints; the lines have the same FWHM as the region A and the amplitude ratio of the \ion{Ca}{II} NIR triplet being emitted from region B is some multiple of the region A. Region B represents this blue-shifted material seen in H$\alpha$. The second scenario has an additional Gaussian representing $\ion{O}{i}~\lambda8446$ fitted independently to a single \ion{Ca}{II} emitting region.

As shown in Fig.~\ref{fig:ca_fitting_comparision}, both scenarios give an acceptable fit to spectrum at $+345$~d. Fitting a single Gaussian emission line representing $\ion{O}{I}~\lambda8446$ gives a reasonable fit with FWHM$\sim$~4000~\kms redshifted by $\sim$~800~\kms. Alternatively, adding an additional \ion{Ca}{II} emission profile we find a good fit at FWHM$\sim$~2000~\kms and blue-shifted by $\sim-2800$~\kms. Although the scenarios are inconclusive, this does not exclude a complex asymmetrical CSM structure producing these multiple emitting regions along the line-of-sight. 

Although both scenarios give reasonable fits, the FWHM and velocities deduced for both scenarios are not seen elsewhere in the spectrum at $+345$~d. It is possible that the region(s) producing the \ion{Ca}{II} NIR triplet is separated from H emitting areas although detailed modelling is needed to confirm. We note however one should expect a similar flux from $\ion{O}{i}~\lambda7774$ when assuming the presence of $\ion{O}{i}~\lambda8446$ which is not the case here. If both lines are produced by recombination, we expect similar relative intensities \citep{NIST_ASD}. Interestingly, this is also trend is also seen in \ip \citep{Graham2014}. 

Our final spectra on $+385$~d and $+420$~d show the \ion{Ca}{II} NIR triplet and [\ion{Ca}{II}] having a broadened appearance compared to earlier spectra. This may indicate an increase in the velocity of the region where these lines form, similar to what is seen in H$\alpha$ in Sect.~\ref{sec:balmer_evo}.

\subsection{Iron Lines}\label{sec:fe}

As temperatures and opacities drop the spectra of many CCSNe become dominated by iron lines, as well as \ion{Na}{i} and \ion{Ca}{II}. We notice persistent permitted Fe group transitions throughout the evolution of \jbu which is likely pre-existing iron in the progenitor envelope. Our initial spectra display the $\ion{Fe}{II} ~\lambda\lambda\lambda~4924,5018,5169$ (multiplet 42) as P Cygni profiles, see Fig.~\ref{fig:dupont_spectra}. At $-31$~d we measure the absorption minimum of \ion{Fe}{II} multiplet 42 at $-750$~\kms. This is the same velocity as the fitted absorption profile from H$\alpha$/H$\beta$ see Fig.~\ref{fig:halpha_spec_evo}A. We can assume that this lines originate in similar regions. 

\begin{figure*}
\centering
\includegraphics[width=\textwidth]{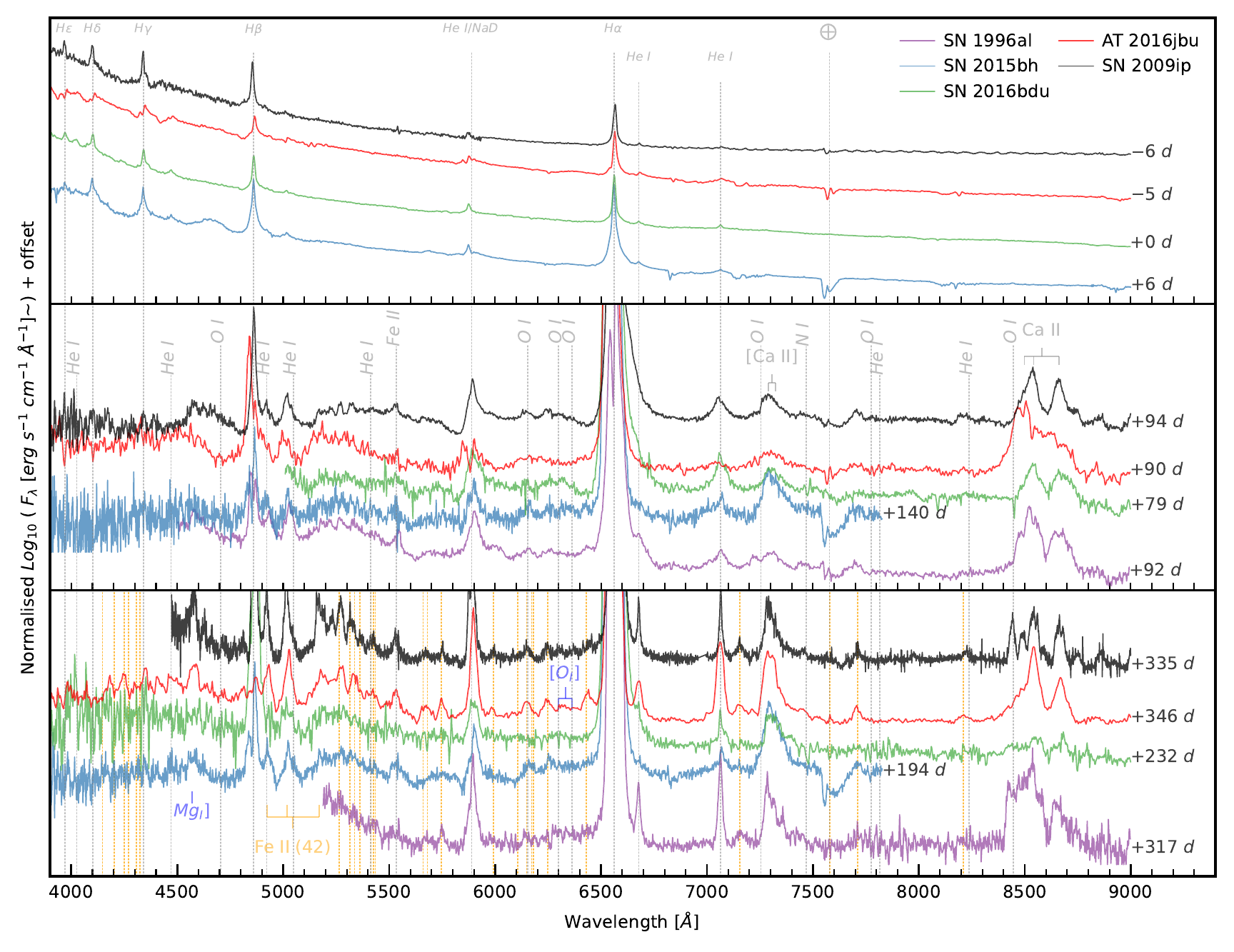}
\caption{ Spectral comparison of \ip-like transients around \textit{Event B} peak (top), three months after \textit{Event B} (middle) and late time spectra around one year later (bottom). We include several strong \ion{Fe}{II} emission lines in the bottom panel as orange vertical lines. We note the remarkable similarities between \jbu and other \ip-like transients at late times.} 
\label{fig:compare_spectra_linear}
\end{figure*}

The \ion{Fe}{II} multiplet 42 appears in our late time spectra, see Fig.~\ref{fig:compare_spectra_linear}. \ion{Fe}{ii} lines in general appear with P Cygni profiles at late times. It is difficult to measure the absorption minimum of the \ion{Fe}{ii} profile due to severe blending. However, using several relatively isolated \ion{Fe}{ii} lines at $+345$~d we measure an absorption minimum of $\sim~-1300$~\kms. The values is similar to the velocity offset for the red and blue emission components seen in H$\alpha$. This suggest that these lines are originating in the same region.

\subsection{Helium Evolution}\label{sec:he}


\begin{figure}
\centering
\includegraphics[width=\columnwidth]{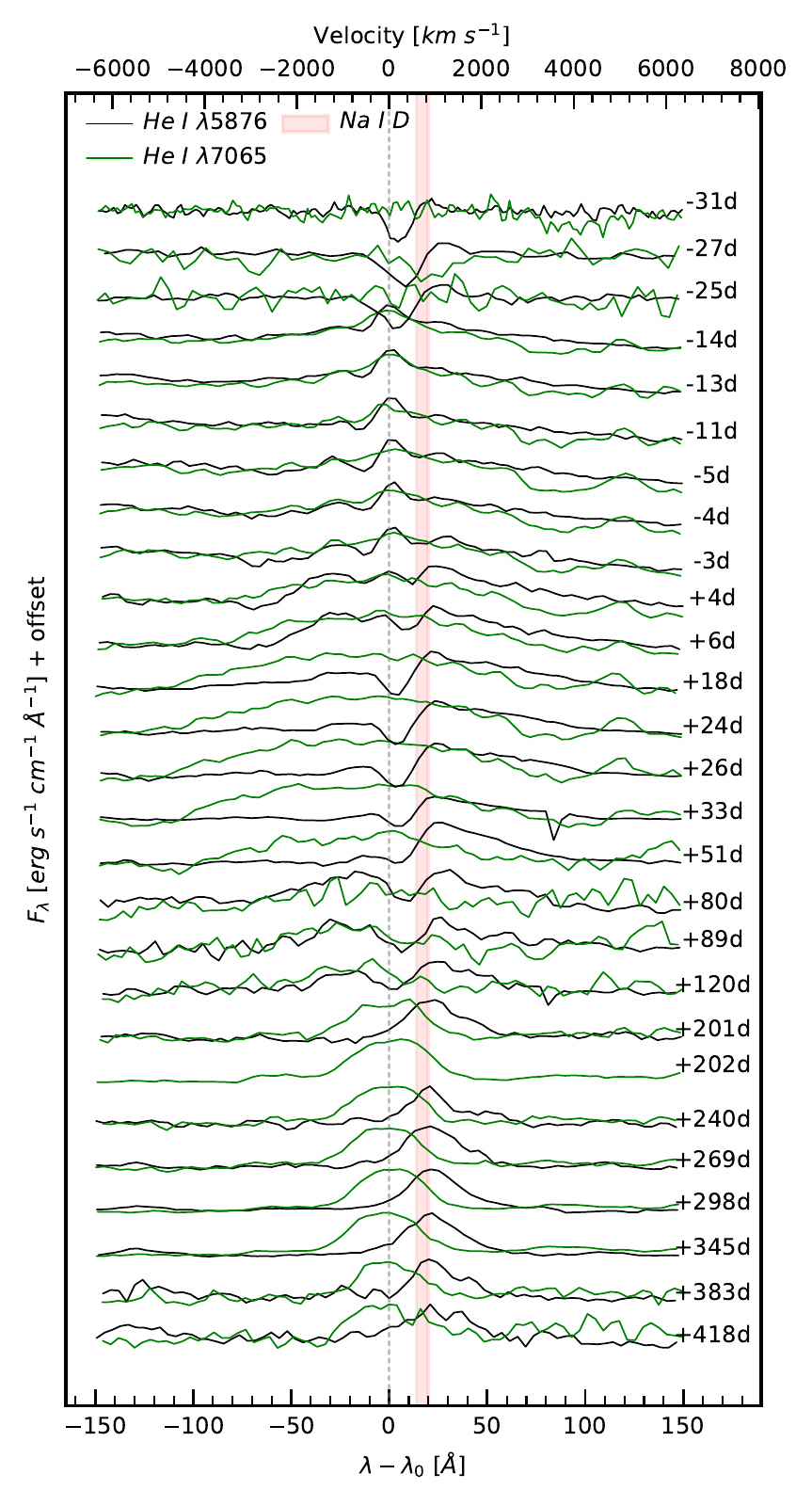}
\caption{Evolution of $\ion{He}{I}~\lambda5876$ (black) and $\ion{He}{I}~\lambda7065$ (green) from NTT+EFOSC2 and DuPont spectra. The rest wavelength of the $\ion{He}{I}$ lines (5876~\AA~ and 7065~\AA) are marked with a vertical line, while Na~I~D$ ~\lambda\lambda~5890,~5895$, is shown by the red vertical band. A velocity scale for the $\ion{He}{I}$ lines is given in the upper axis. Each spectrum has been normalised to a peak value of unity.}
\label{fig:helium_evo}
\end{figure}

None of the \ion{He}{I} lines display the degree of asymmetry seen in hydrogen. Transients exist displaying double-peaked helium lines, such as the Type Ibn SN~2006jc; \citep{Foley2007,Pastorello2008}, as well as some displaying asymmetric \ion{He}{I} and symmetric H emission e.g. the Type Ibn/IIb SN~2018gjx \citep{Prentice_2020}.

We show the evolution of $\ion{He}{I} ~\lambda5876$ (black line) and $\ion{He}{I} ~\lambda7065$ (green line)in Fig.~\ref{fig:helium_evo}.  $\ion{He}{I} ~\lambda7065$ first appears in emission on $-14$~d with a boxy profile that is poorly fit with a single Lorentzian emission line. $\ion{He}{I} ~\lambda7065$ then becomes more symmetric by $+18$~d. Note the blue absorption feature in H$\alpha$ is also first seen at this time. The line begins to broaden over the next month, peaking at FWHM$\sim3400$~\kms at $\sim~+28$~d. After $+51$~d, $\ion{He}{I} ~\lambda7065$ is no longer detected with any reasonable S/N.

Interestingly, $\ion{He}{I} ~\lambda7065$ then re-emerges at +200~d, the emission feature has FWHM $\sim1100$~\kms centered at rest wavelength. We see this same FWHM in the red and blue shoulders in H$\alpha$ (Sect.~\ref{sec:balmer_evo}). We find that a single emission profile matches the $\ion{He}{I} ~\lambda7065$ line well after +200~d. However, motivated by the multi-component profile of H$\alpha$ we also find that $\ion{He}{I}~\lambda7065$ after +200~d can be fitted equally well with two emission components. In this case, both components are offset by $\sim\pm 400$~\kms from their rest wavelength, and each has a FWHM of $\sim1000$~\kms. Unlike H$\alpha$, no third core emission component is needed. 

For $\ion{He}{I} ~\lambda5876$, in our $-31$~d spectrum there is a clear P Cygni profile centered at 5898~\AA. The emission is likely caused by Na~I~D with the possibility of some absorption contamination from $\ion{He}{I} ~\lambda5876$. We measure a velocity offset of $\sim-450$~\kms with respect to 5890~\AA. At $-13$~d, $\ion{He}{I} ~\lambda~5876$ emerges and has a complicated, multi-component profile with contamination from Na~I~D. Emission centered on 5876~\AA\ persists until +20~d, after which the emission returns to being dominated by Na~I~D.

Low resolution spectra preclude further investigation, but if $\ion{He}{I}~\lambda7065$ is composed of two emission profiles, these two emission regions are at significantly lower velocity when compared to the similar components in H$\alpha$. An increase in the strength of \ion{He}{I} was also seen in the Type IIn \al and was interpreted as a signature of strengthening CSM interaction \citep{Benetti_2016}. 

$\ion{He}{I} ~\lambda6678$ evolves in a similar manner to $\ion{He}{I} ~\lambda7065$, but shows a clear P Cygni profile as early as $-14$~d with an absorption trough at $\sim-500$~\kms, similar to H$\alpha$. After the seasonal gap $\ion{He}{I}~\lambda5876$ is not clearly seen. At +345~d we measure a Gaussian emission profile centered at 5897~\AA\ with a FWHM $\sim1800$~\kms. This is likely dominated by Na~I~D with minor contamination from $\ion{He}{I}~\lambda5876$. The FWHM value for this line suggests that it is coming from the site of \jbu and not due to host contamination.

\begin{figure}
\includegraphics[width = \columnwidth]{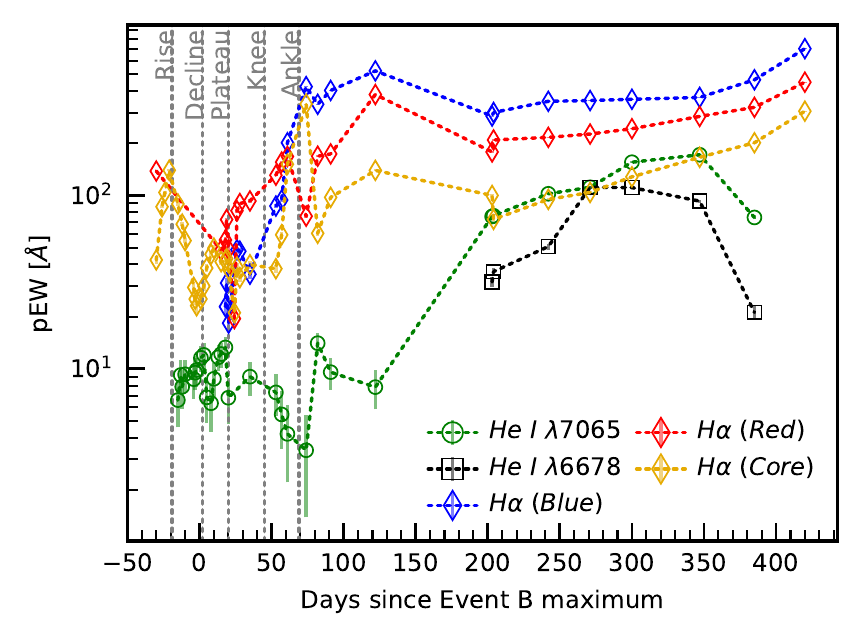}
\caption{Evolution of pEW for $\ion{He}{i}~\lambda~6678, \lambda~7065$ and H$\alpha$ components. The \ion{He}{I} emission appears to be roughly constant until the \textit{knee}/\textit{ankle} stage when it increases rapidly. After $\sim~+300$~d the pEW of \ion{He}{i} again begins to decrease. The measurement of pEW is based on a single emission component fit which provides a reasonable fit at late times. $\ion{He}{i}~\lambda~6678$ is not plotted for $t<220$~d due to its low pEW and contamination from H$\alpha$.}
\label{fig:he_EW}
\end{figure}

We plot the evolution of the pseudo-Equivalent Width (pEW) (a pseudo-continuum is fitted over a small wavelength window) of the two seemingly isolated $\ion{He}{i}~\lambda\lambda~6678,7065$ emission lines in Fig.~\ref{fig:he_EW}.  We note that there is little change in pEW for the first $\sim 120$ days. After the seasonal gap, both emission lines increase dramatically in pEW, until $\sim~+300$~d after which the pEW declines. A similar jump in \ion{He}{I} was seen in \al \citep{Benetti_2016}. This decline coincides with the narrowing and increase in amplitude of the blue, red, and core emission components of H$\alpha$. 

\ion{He}{I} emission is expected to be formed in the de-excitation/recombination region of the shock wave \citep{Chevalier1978,Gillet_2014}. As mentioned in Sect.~\ref{sec:balmer_evo}, after $\sim$~2 months, the blue shifted emission in H$\alpha$ grows in amplitude and narrows considerably, likely due to changing opacities. This jump in pEW may represent a time when shocked material is no longer obscured and photons can escape freely from the interaction sites. We reach a similar conclusion for \ion{He}{i}. If the trend in both \ion{He}{i} lines is linked to the H$\alpha$ emitting regions, then it is likely that the late time \ion{He}{i} might also be double-peaked.

Fig.~\ref{fig:lc} shows a rebrightening/flattening after the seasonal gap. This is seen best in {\it Gaia}-\textit{G}. The trend seen in \ion{He}{i} $\lambda~6678$ and $\lambda~7065$ pEW may follow the interaction of the shock front with some clumpy dense material far away from the progenitor site. This would reflect a stratified CSM profile possibly produced by the historic eruptions, or possibly a variable wind, in \jbu. 

\subsection{Forbidden Emission Lines}\label{sec:forbiddenlines}

A clear sign of a terminal explosion is forbidden emission lines from material formed during explosive nucleosynthesis/late-time stellar evolution. All CCSNe will eventually cool down sufficiently for the photosphere to recede to the innermost layers of the explosion. We expect to see the signatures of material synthesised in the explosion as well as material produced in the late-stages of stellar evolution such as $[\ion{O}{I}]~\lambda\lambda~6300, 6364$ or $\ion{Mg}{I}]~\lambda~4571$ \citep{Jerkstrand2017}.
%
%
Fig.~\ref{fig:compare_spectra_linear} shows the late time spectra of \jbu highlighting prominent emission lines. Tenuous detections are made of [\ion{O}{I}] and \ion{Mg}{I}], although these lines are much weaker than are typically seen during the nebular phase of CCSNe. 
Late time spectra show that there is on-going CSM interaction for \jbu, as is clear for the double-peaked H$\alpha$ emission. The spectra are still relatively blue (i.e. Fig.~\ref{fig:compare_spectra_linear}, $\lambda\lesssim5600$~\AA) even after 1~year, again indicating interaction in the CS environment.

It is a common conclusion for \ip-like transients that there are only tenuous signs of core-collapse \citep{Fraser13,Benetti_2016}. \citet{Fraser13} find no clear signs of any such material during the late time nebular phase of \ip. \ip\ showed little indication of a nebular phase and in 2012 showed spectral features similar to its 2009 appearance. \citet{Benetti_2016} finds no evidence of nebular emission features in SN~1996al even after 15~yrs of observations. For \jbu\ one may posit that if the transient is indeed a CCSNe, on-going interaction has led to densities too high for forbidden lines to form. Alternatively, fallback onto a compact remnant could result in an apparently small mass of synthesized heavy elements, and hence an absence of nebular CCSN features. We will expand further on the nature of \jbu\ and \ip-like transients, their powering mechanism and the possibility that the progenitor survived, in \paperII.

\subsection{UV spectrum}\label{sec:UVspectra}

\begin{figure*}
\centering
\includegraphics[width=\textwidth]{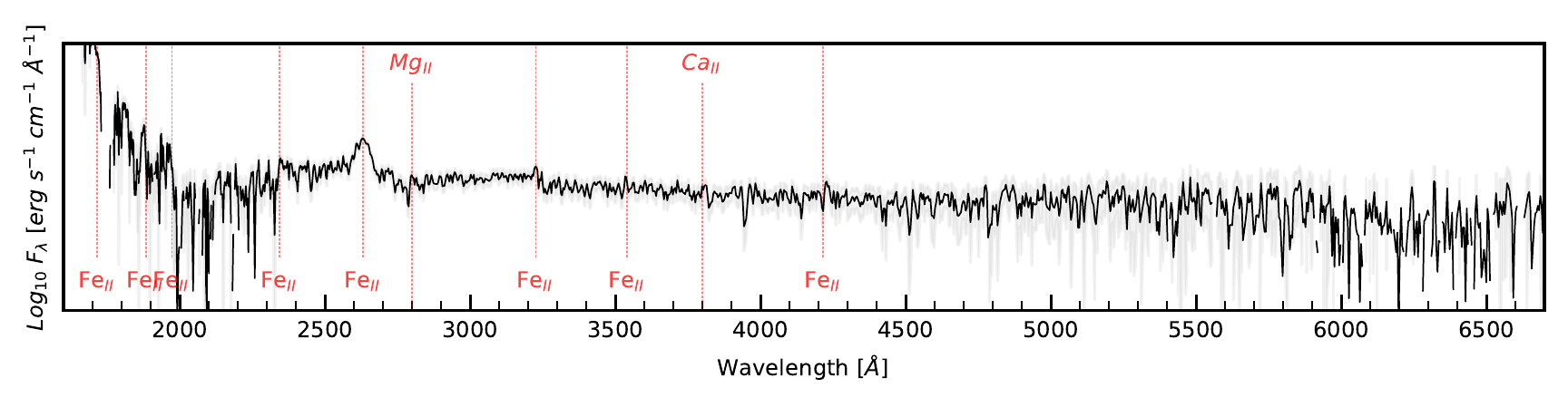}
\caption{{\it Swift} + UVOT spectrum for \jbu taken on 2017 January 22 (MJD: 57775, Phase: $-18$~d). Wavelength in given in rest frame and the spectrum is corrected for Galactic extinction ($A_V$ = 0.556~mag). The spectrum is given in black with the grey shaded region showing the uncertainty. The trough around 2000\AA\ is likely noise which is likely exacerbated by our extinciton correction.}
\label{fig:swift_spectra}
\end{figure*}

We present a single UV spectrum in Fig.~\ref{fig:swift_spectra} taken with {\it Swift}+UVOT on 2017 January 22. The spectrum has quite low S/N towards the red with a very tenuous detection of the Balmer series. It is likely that $\lambda~>~4000$~\AA~ is affected by second order contamination. The continuum of \jbu deviates significantly from a blackbody at short wavelengths ($\lambda < 2400$~\AA) mainly due to blends of lines of singly ionized iron-peak elements.

A broad (FWHM $\sim$~5000~\kms) emission line is the strongest feature seen. It is centered at $\sim 2630$~\AA\ and is well fitted with a single Gaussian. We are unsure of the identification of this emission line, however there is a strong $\ion{Fe}{II}$ line at $\sim$ 2631~\AA\ \citep{NIST_ASD,Nave_1994}. 

It is curious that there is a strong $\ion{Fe}{II}$ line here and no other emission features at comparable strength. \textit{Swift} observations of \ip do show this emission line but it is much weaker than that seen in \jbu \citep{Margutti2014}. This particular emission line has been seen in several Type IIP SNe with UV coverage such as SN~1999em and SN~2005cs \cite[see][and references therein]{Gal_Yam_2008}. However, the Type IIP SNe discussed by \citet{Gal_Yam_2008} also show strong emission from $\ion{Mg}{II}~\lambda~2800$. \jbu shows a weaker P Cygni feature centered at 2800~\AA\ with an absorption at $\sim-1200$~\kms which is likely due to $\ion{Mg}{II}~\lambda~2800$.
Detailed spectral modelling is needed to secure this line identification.

\subsection{NIR spectra}\label{sec:NIRspectra}

\begin{figure*}
\centering
\includegraphics[width=\textwidth]{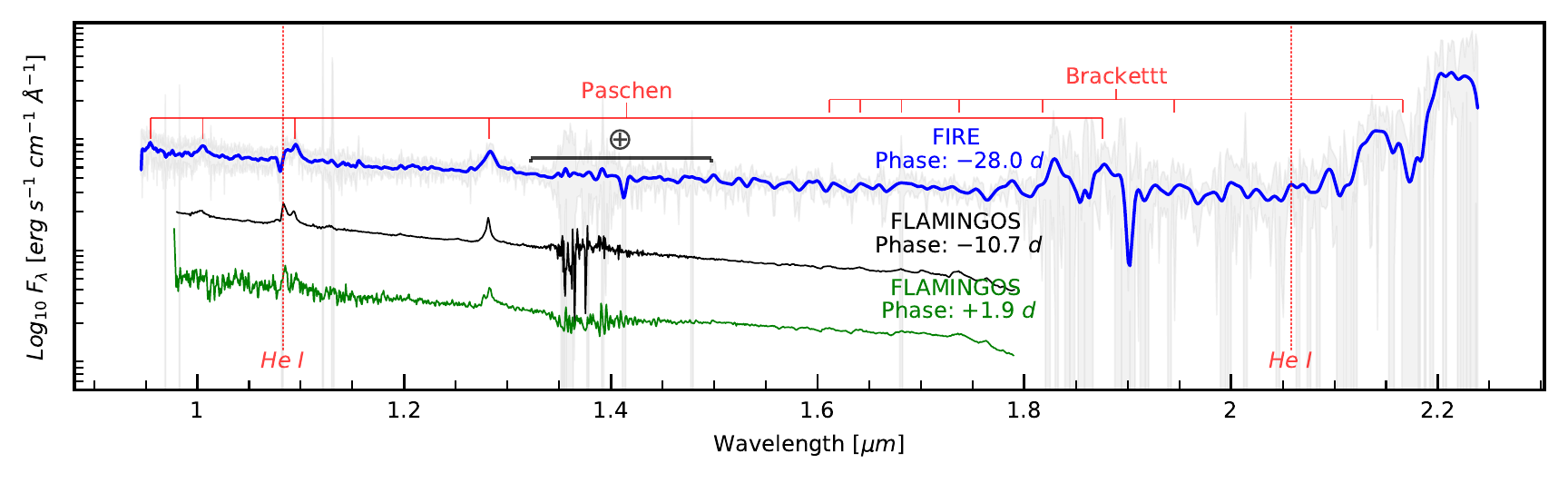}
\caption{NIR spectra of \jbu, covering the peak of \textit{Event A} as well as the rise and peak of \textit{ Event B} . H and \ion{He}{I} are clearly seen in all spectra. The FIRE spectrum (blue) has been smoothed for presentation and shows what appears to be an excess redwards of $2.05\mu m$. This excess is likely due to spectra being saturated by the bright $K$-band sky.}
\label{fig:IR_spectra}
\end{figure*}

We present our NIR spectra in Fig.~\ref{fig:IR_spectra} covering the peak of \textit{Event A} as well as the rise and peak of \textit{ Event B}. Pa$\beta~\lambda~12822$ follows the same evolution as H$\alpha$, with a strong blue absorption profile that is not present in the $-31$~d FIRE spectrum but which appears in the FLAMINGOS-2 $-12$~d spectra. At this phase the blue absorption is already seen in H$\alpha$ and H$\beta$. Pa$\beta$ is also broader at $-31$~d and narrows at $-12$~d, similar to the H$\alpha$ evolution shown in Fig.~\ref{fig:halpha_spec_evo} at $-31$~d and $+1$~d.

There is a strong \ion{He}{I} $\lambda10830$ line blended with Pa$\gamma$. At $-31$~d this line appears in absorption at rest wavelength, while by $-12$~d the line is in emission. This helium feature may be thermally excited and this is supported by the blackbody temperature seen peaking at this time (see \paperII). We see an absorption trough bluewards of $\lambda10830$ which may be associated with Pa$\gamma~\lambda~10941$ (as a similar absorption is seen in Pa$\beta$). There appears to be a flux excess beyond 2.1~$\upmu$m in the FIRE spectrum at $-31$~d. This may represent emission from a CO bandhead, possibly signifying some pre-existing dust during \textit{Event A}. However, the S/N is extremely low in this region of the spectrum (see the grey shaded region in Fig.~\ref{fig:IR_spectra}), and it is likely that the apparent ``excess'' is due to bright $K$-band sky contamination rather than CO emission.

\section{Discussion}\label{sec:dicussion}

We will discuss \jbu and their relation to \ip-like objects, mainly their photometric similarities in Sect.~\ref{sec:phot_comp} and their spectroscopic appearance in Sect.~\ref{sec:spec_comp}, in particular the appearance of their H$\alpha$ emission profiles is varies times during their evolution (Sect.~\ref{sec:discuss_halpha}).

\subsection{\jbu and other \ip-like transients}\label{sec:09ip-like} 

For this paper we focus the discussion on the photometric and spectral comparison between \jbu and similar transients. In \paperII we discuss topics including the progenitor of \jbu using pre-explosion images, the environment around the progenitor and a non-terminal explosion scenario.

\subsubsection{Photometric Comparison}\label{sec:phot_comp}

\begin{figure*}
\includegraphics[width=\textwidth]{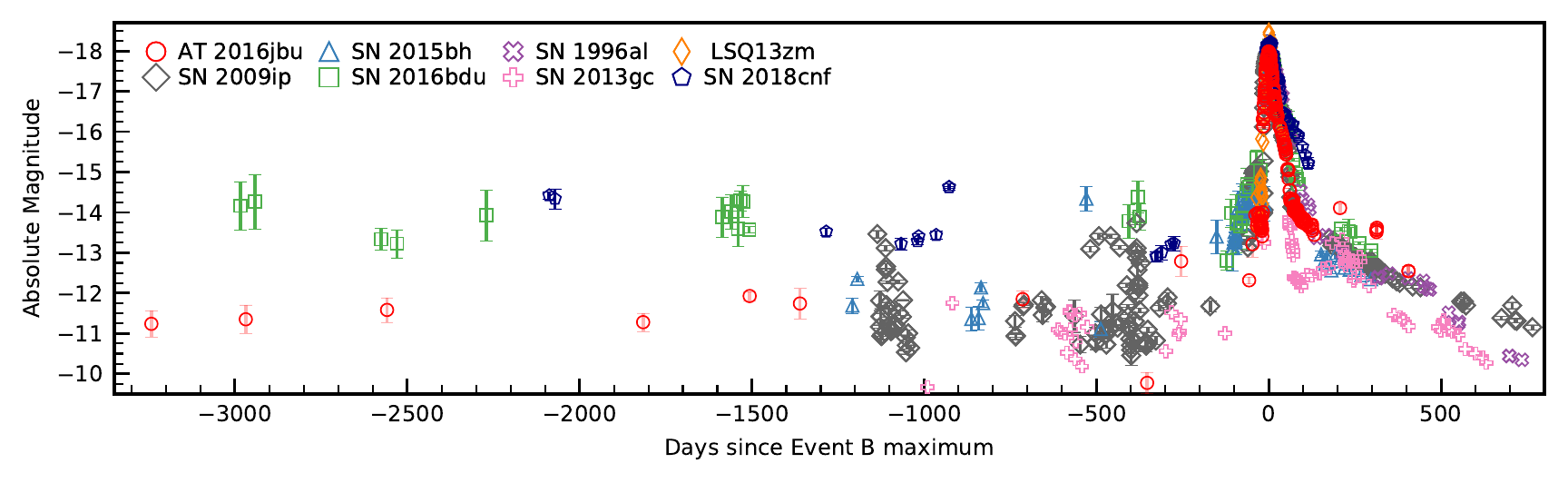}
\caption{Pre-explosion outbursts and the main luminous event for the sample of \ip-like transients. \ip (Sloan r) is taken from \protect\citet{Fraser13, Graham2014}, \bh (\textit{R}) from \protect\citet{Thone2017}, \bdu (\textit{r}) from \protect\citet{Pastorello_2017}, \gc (R)  from \protect\citet{Reguitti2018}, \al (\textit{\textit{R}}) from \protect\citet{Benetti_2016}, \cnf (\textit{\textit{r}}) from \protect\citet{Pastorello2019}, and \lsq (\textit{R}) is taken from \protect\citet{Tartaglia2016}. All data given in Vega magnitudes \citep{Blanton_2007}. We do not show limiting magnitudes in this figure for clarity. All events show an initial rise to a magnitude of $\sim$-14 (if coverage available) followed by a second rise to $\sim-18$ roughly 30 days later. Our sample of \ip-like transients all show outbursts in the months/years prior to their luminous events.}
\label{fig:compare_lc}
\end{figure*}

We compare the $R$/$r$-band lightcurves of a sample of \ip-like transients events in Fig.~\ref{fig:compare_lc}. In cases where \textit{r}-band photometry was not available, Johnson-Cousin \textit{R}-band is shown. The adopted extinction and distance moduli are given in Table \ref{tab:objects}. The photometric evolution for \ip-like transients is undoubtedly similar. Our sample of transients all show a series of outbursts in the years prior to \textit{Event A}, as seen in Fig.~\ref{fig:compare_lc}. This has been described as historic ``\textit{flickering}'' by \citetalias{Kilpatrick2018}. \jbu shows several clear detections within $\sim$~10 years before the peak of \textit{Event B}. Similar outbursts are seen in other \ip-like transients (see Fig.~\ref{fig:compare_lc}). 

The duration of \textit{Event A} varies between each transient. For \ip, \textit{Event A} lasts for $\sim$~1.5 months \citep{Fraser13} and rises to $\sim-15$~mag. \lsq shows a rise to $\sim-14.8$~mag and has a time frame of a few weeks \citep{Tartaglia2016}. All transients show a fast rise of $\sim$~17 days to maximum in \textit{Event B} to $\sim-18\pm0.5$ mag followed by a rapid/bumpy decay. \citet{Kiewe12} found that a magnitude of $-18.4$ is typical for Type IIn SNe. Using a larger sample size, \citet{Nyholm_2020} find a larger value for the mean value although \textit{Event B} peak is still within a standard deviation of this. 

\begin{figure}
\centering
\includegraphics[width=\columnwidth]{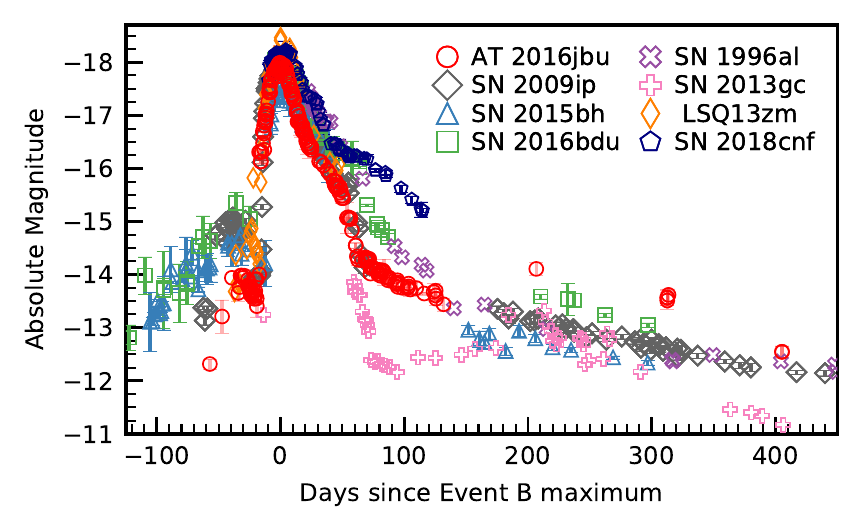}
\caption{Same as Fig.~\ref{fig:compare_lc}, but focusing \textit{Event A/B}. All \ip-like transients show a similar \textit{Event B} (lightcurve), although \textit{Event A} tends to be more diverse (if observations are available). \jbu shows a major rebrightening after $\sim200$~d days not seen in other \ip-like transients.}
\label{fig:compare_lc_zoom}
\end{figure}

Curiously, several of the transients in our sample show their first initial bump around the same time, approximately 20 days post maximum; see Fig.~\ref{fig:compare_lc_zoom}. \jbu shows no major bumps in its lightcurve, but instead flattens slightly, whereas \ip and \cnf show a clear and prominent bump at $\sim20$~d.

From $\sim$~60 -- 120~d, \jbu appears to follow the extrapolated decline of \ip (see Fig.~\ref{fig:compare_lc_zoom}). However, when \jbu emerged from behind the Sun at +200~d, it shows a large increase in magnitude in all bands. No other \ip-like transient shows a comparable behaviour. At $\sim200$~d, \jbu is almost 1~mag brighter than \ip. We see a change in \ion{He}{I} pEW  (see Sect.~\ref{sec:he}) which is not clearly seen in H$\alpha$ at this time and may reflect enhanced interaction with a complex CSM environment. 

\subsubsection{Spectroscopic Comparison}\label{sec:spec_comp}

The spectra of \ip-like transients remain remarkably similar as they evolve. Fig.~\ref{fig:compare_spectra_linear} shows our sample of extinction corrected \ip-like transients at several phases during their evolution. All objects initially appear similar to Type IIn SNe, with $T_{BB} \sim 10,000~K$ and prominent narrow lines seen in the Balmer series. 


\begin{figure}
\centering
\includegraphics[width=\columnwidth]{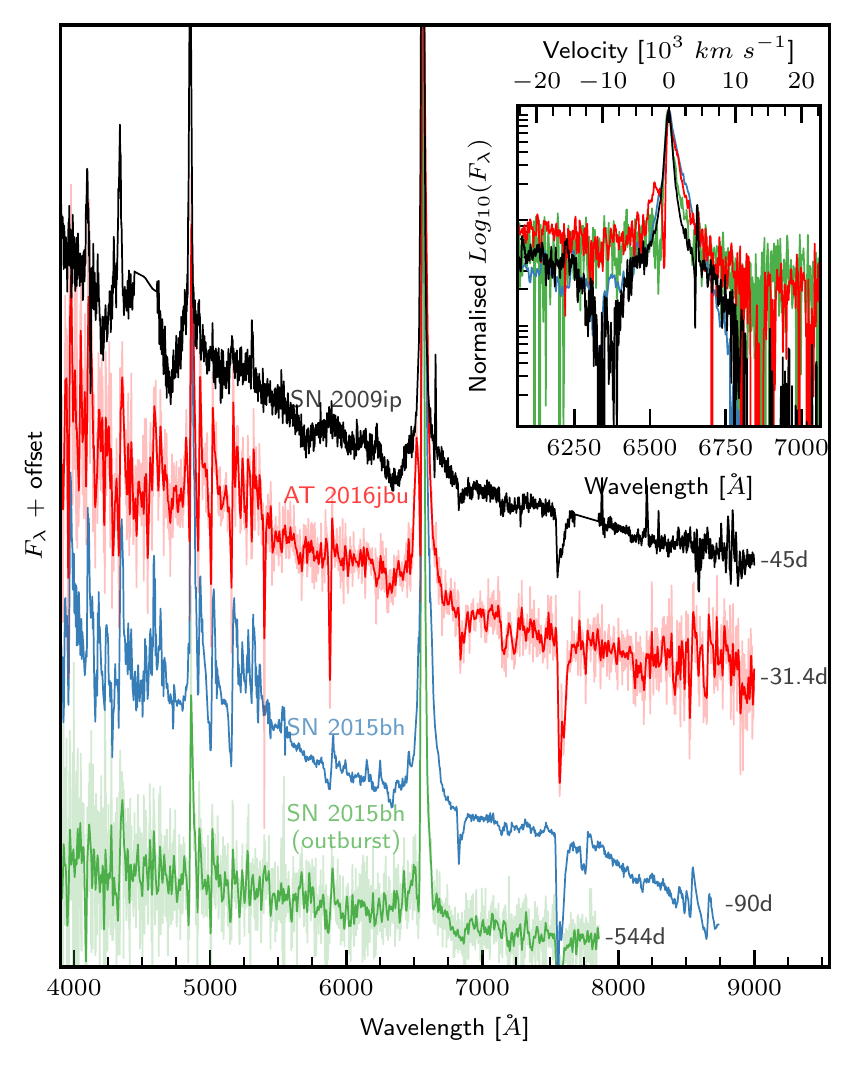}
\caption{ Spectral comparison of \ip, \jbu,and \bh during their respective A events. Also included is the spectrum of \bh during an apparent LBV outburst in 2013 \citep{Thone2017}. The inset shows a close up of H$\alpha$, normalised to the emission peak to highlight the velocity structure on \ip. \bh has been shifted bluewards by 2~$\si{\angstrom}$ to match the other H$\alpha$ lines. \jbu and \bh have been smoothed with a Gaussian kernel for clarity. }
\label{fig:EventA_spectra}
\end{figure}


In Fig. \ref{fig:EventA_spectra} we compare the appearance of \ip, \jbu,and \bh around the time of their Event A maxima. We also include the apparent pre-explosion outburst of \bh \citep{Thone2017} seen in 2013 ($\sim$1.5 years before the possible SN). This spectrum of \bh shows a very narrow H$\alpha$ profile that is fitted well with a single P-Cygni profile, and is reminiscent of a LBV in quiescence \citep{Thone2017}. All four transients shows a blue continuum with narrow emission features seen mainly in the Balmer series and Fe. Where they differ is in the presence or absence of a {\it broad} component in H$\alpha$. \ip is dominated by a $\sim13000$~\kms absorption feature and strong narrow emission line. \jbu shows a broader emission component ($FWHM~\sim~2600$~\kms) with a P-cygni absorption feature at $\sim~-700$~\kms. Similarly \bh shows a broad emission profile like \jbu and also lacks any broad absorption at this time. Although these transients evolve similarly (see below), our earliest \textit{Event A} spectra suggest that the explosion mechanism for these transients may be quite diverse. This argument is strengthened by the variety among \textit{Event A} lightcurves (inset in Fig. \ref{fig:compare_lc_zoom}). It is a puzzle why these transients appear to evolve similarly during and after \textit{Event B} but show such diversity during \textit{Event A}. In particular, the presence of fast material during \textit{Event A} of \ip was suggested to be evidence that the progenitor has undergone core-collapse \citep{Mauerhan2013}. If this is true, then the absence of high velocity features in the other transients must be explained by different CSM configuration or viewing angle effects. If geometry is a strong contributor to the appearance of these transients, then one can not ignore the possibility that \textit{Event A} for each transient is a result of a similar explosion mechanism e.g. a low luminosity Type II SN \citep{Elisa-Rosa2016,Mauerhan2013,Margutti2014}.

\subsubsection{H$\alpha$ comparison}\label{sec:discuss_halpha}

\begin{figure*}
\centering
\includegraphics[width=\textwidth]{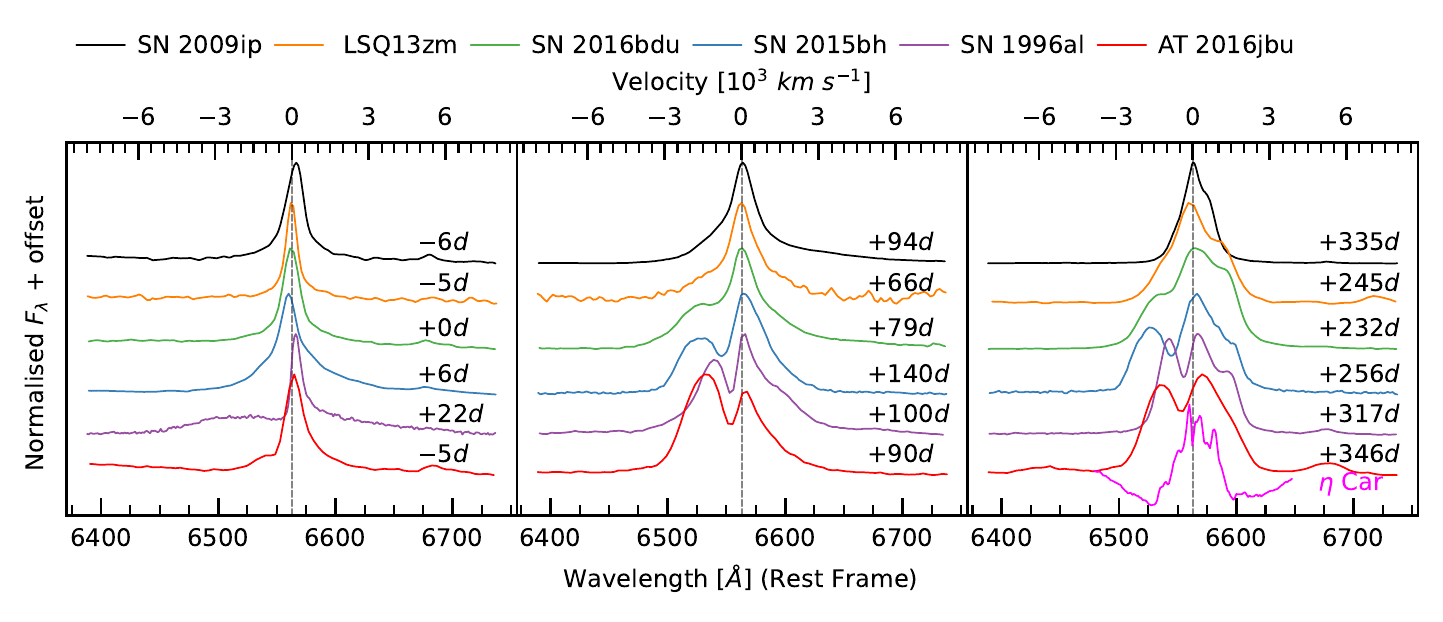}
\caption{H$\alpha$ spectral comparison between \ip-like transients. Spectra are plotted after normalising with respect to the peak of H$\alpha$, with arbitrary flux offsets for clarity. Spectra were de-reddened using the parameters given in Table. \ref{tab:objects}. Early time spectra show a Type IIn SNe-like profile with narrow emission. while spectra $\sim$~3 months later show the emergence of a blue and red shoulder in each profile. At late times, H$\alpha$ forms a double-peaked emission profile, aside from in the case of \ip (although here there is still evidence for a red shoulder component). The difference in line shape is most likely due to inclination, an idea we elaborate on in Sect.~\ref{sec:dicussion}. We also show the spectrum of \etacar (at $\sim+150$~yr)}
\label{fig:compare_spectra_halpha}
\end{figure*}

We show a zoom in on H$\alpha$ in Fig.~\ref{fig:compare_spectra_halpha}, where the spectra are plotted in order of ``\textit{double-peaked}''-ness i.e according to the level of double-peaked nature of the H$\alpha$ line profile. We arbitrarily define \textit{double-peaked}-ness as the strength and separation between the two emission peaks (if any) seen in H$\alpha$. All objects also appear to show an additional high velocity blue absorption in their Balmer lines (panel B of Fig.~\ref{fig:compare_spectra_halpha})\footnote{The spectroscopic data for \al only begins at 22 days past \textit{Event B}, when we can already see the emergence of a broad blue component.}.  At intermediate times, $\sim$~3 months after maximum, all transients (excluding \ip)  show clear evidence of strong multi-component profiles. \jbu shows the strongest appearance of a double-peaked profile, whereas \ip show the least, with weak evidence of some blue excess.

\begin{figure}
\centering
\includegraphics[width=\columnwidth]{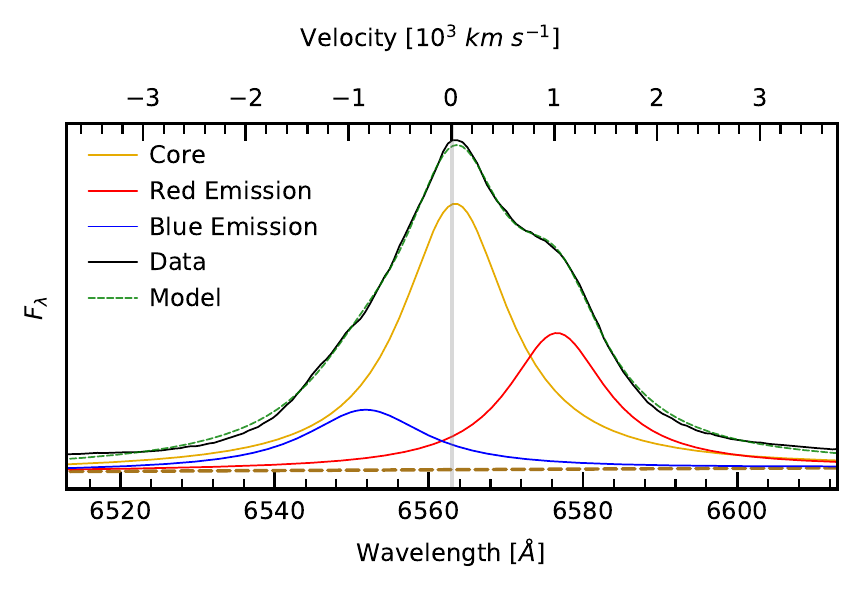}
\caption{Spectral decomposition of the H$\alpha$ profile for \ip at $+335$~d. Spectra from the DEep Imaging Multi-Object Spectrograph (DEIMOS; \citealp{faber2003}), was fitted as mentioned in Fig.~\ref{fig:halpha_spec_evo}. A three-component model reproduces the observed H$\alpha$ profile at late times.}
\label{fig:ip_halpha_fit} 
\end{figure}

After $\sim$~10 months, all transients show multi-component profiles in H$\alpha$. Each transient displays different velocity and FWHM values for their red and blue components. For \ip, \citet{Fraser2015} notes a red component at +500~\kms at late times, this shoulder is also seen in H$\beta$. We measure the same component at $+625$~\kms while fitting for an additional blue component at $-510$~\kms. Our fit is illustrated in Fig.~\ref{fig:ip_halpha_fit}. In the case of \ip, this red shoulder only appeared at $\sim$~5.5 months after maximum light, whereas there is evidence of this red shoulder as early as a week after maximum for \jbu. This is likely due to geometric inclination effects along the line-of-sight, with \ip being the most edge on and \jbu being the more face on. Ejecta-disk models by \citet{Kurfurst2020} show this profile shape versus line-of-sight effect. 

We include a close up of the H$\alpha$ profile of \etacar in Fig.~\ref{fig:compare_spectra_halpha}, based on VLT+MUSE observations taken on 2014 Nov. 13. This spectrum was extracted from spaxels with a 14\arcsec\ radius of \etacar after masking nearby stars. \etacar displays a multi-peaked H$\alpha$ profile similar to what we see in our \ip-like transients events, albeit at a lower velocity. A similarly shaped profile is also seen in spectra obtained from light echoes of the \textit{Great Eruption} (GE) \citep{Smith_2018}. This resemblance raises the tantalising possibility that \etacar and \ip-like transients share similar progenitors or progenitor systems.

To date, it is still uncertain what caused the GE in \etacar, although commonly discussed scenarios include a major eruption triggered by a merging event in a triple stellar system \citep{Smith_2018}, mass transfer from a secondary star during periastron passages \citep{Kashi_2010} or even a pulsational pair-instability explosion \citep{Woosley_2007}. 

Despite the asymmetric H$\alpha$ emission lines, curiously no other lines show such asymmetry, in particular \ion{He}{I}. However, we cannot exclude that this is simply due to lower S/N in these other lines, or that their lower velocities mean that any signs of asymmetry are masked by our moderate instrumental resolution.

\section{Conclusion}\label{sec:conclusions}

In this paper, we have presented the results of our follow-up campaign for \jbu consisting of high cadence photometry up to $\sim$~1.5 years after maximum light, together with spectra spanning $-31$ to $+420$~days covering the UV, optical and NIR. We also present historical observations over the preceding decade from ground based observations.

In summary, the salient points of this work are:

\begin{itemize}[leftmargin=*,labelindent=1em]

 \item \jbu displays variability in the years prior to maximum light, with outbursts reaching $M_r\sim-11.5$~mag, and a double-peaked lightcurve. The first peak reaches $M_r\sim-13.5$~mag and the second reaches a SN-like magnitude of $M_r\sim-18.26$~mag, with both peaks separated by $\sim$~1~month.
 \item \jbu shows a smooth lightcurve with a major re-brightening event occurring after the seasonal gap ($\sim$~200~days). An increase in \ion{He}{I} emission is seen during this time, which may be a sign of increased interaction.
 
 \item \jbu appears spectroscopically and photometrically alike to \ip, \bh, \bdu, \al, \gc and \cnf. However, the increase in brightness at $\sim+200$~d is unique to \jbu with respect to our sample of \ip-like transients. The color evolution is similar amongst all \ip-like transients. Color changes can be linked with the appearance of the red and blue emission components seen in H$\alpha$.
 
 \item The H$\alpha$ profiles of each transient show an apparent continuum of asymmetry and we deduce that this may be caused by an geometric inclination effect. 

 \item \jbu and other \ip-like transients do not exhibit signs of explosive nucleosynthesis at late times such as $[\ion{O}{I}]~\lambda\lambda~6300, 6364$ or $\ion{Mg}{I}]~\lambda~4571$. On-going CSM interaction may be inhibiting these features and/or obscuring their emitting regions. 
 
\end{itemize}

\jbu and the \ip-like transients are peculiar objects. If they are indeed SNe then their progenitors undergo an unusual and poorly-understood series of eruptions in the years prior to core-collapse. If these events are non-terminal and the progenitor star will be revealed in the future, it begs the question what sort of mechanism can produce such an energetic explosion.

In \paperII we continue the discussion of \jbu and \ip-like transients using the data presented here, focusing on the local environment, the progenitor and modelling of the light-curve.

\section*{Acknowledgements}

S. J. Brennan acknowledges support from Science Foundation Ireland and the Royal Society (RS-EA/3471). M.F is supported by a Royal Society - Science Foundation Ireland University Research Fellowship. T.M.B was funded by the CONICYT PFCHA / DOCTORADOBECAS CHILE/2017-72180113 and acknowledges their financial support from the Spanish Ministerio de Ciencia e Innovaci\'on (MCIN), the Agencia Estatal de Investigaci\'on (AEI) 10.13039/501100011033 under the PID2020-115253GA-I00 HOSTFLOWS project, and from Centro Superior de Investigaciones Cient\'ificas (CSIC) under the PIE project 20215AT016. K.M. is funded by the EU H2020 ERC grant no. 758638. T.W.C acknowledges the EU Funding under Marie Sk\l{}odowska-Curie grant H2020-MSCA-IF-2018-842471, and thanks to Thomas Kr{\"u}hler for GROND data reduction. M.N is supported by a Royal Astronomical Society Research Fellowship. B.J.S is supported by NSF grants AST-1908952, AST-1920392, AST-1911074, and NASA award 80NSSC19K1717. M.S is supported by generous grants from Villum FONDEN (13261,28021) and by a project grant (8021-00170B) from the Independent Research Fund Denmark. L.H acknowledges support for Watcher from Science Foundation Ireland grant 07/RFP/PHYF295. Time domain research by D.J.S. is supported by NSF grants AST-1821987, 1813466, \& 1908972, and by the Heising-Simons Foundation under grant \#2020-1864. N.E.R. acknowledges support from MIUR, PRIN 2017 (grant 20179ZF5KS). L.G. acknowledges financial support from the Spanish Ministry of Science, Innovation and Universities (MICIU) under the 2019 Ram\'on y Cajal program RYC2019-027683 and from the Spanish MICIU project PID2020-115253GA-I00. Support for TW-SH was provided by NASA through the NASA Hubble Fellowship grant HST-HF2-51458.001-A awarded by the Space Telescope Science Institute, which is operated by the Association of Universities for Research in Astronomy, Inc., for NASA, under contract NAS5-26555.Support for JLP is provided in part by ANID through the Fondecyt regular grant 1191038 and through the Millennium Science Initiative grant ICN12\_009, awarded to The Millennium Institute of Astrophysics, MAS.  D.A.H and D.H are supported by AST-1911151, AST19-11225, and NASA Swift grant 80NSSC19K1639. G.P acknowledge support by the Ministry of Economy, Development, and Tourism’s Millennium Science Initiative through grant IC120009, awarded to The Millennium Institute of Astrophysics, MAS. L.T. acknowledges support from MIUR (PRIN 2017 grant 20179ZF5KS). H.K. was funded by the Academy of Finland projects 324504 and 328898.

This research made use of Astropy\footnote{\url{http://www.astropy.org}}, a community-developed core Python package for Astronomy \citep{astropy:2013, astropy:2018}. This research made use of data provided by Astrometry.net\footnote{\url{https://astrometry.net/use.html}}. This research has made use of the NASA/IPAC Extragalactic Database (NED), which is operated by the Jet Propulsion Laboratory, California Institute of Technology, under contract with the National Aeronautics and Space Administration. We acknowledge the use of public data from the Swift data archive. This work made use of data supplied by the UK Swift Science Data Centre at the University of Leicester. We acknowledge Telescope Access Program (TAP) funded by the NAOC, CAS, and the Special Fund for Astronomy from the Ministry of Finance.  This work was partially supported from Polish NCN grants: Harmonia No. 2018/30/M/ST9/00311 and Daina No. 2017/27/L/ST9/03221. Parts of this research were supported by the Australian Research Council Centre of Excellence for All Sky Astrophysics in 3 Dimensions (ASTRO 3D), through project number CE170100013. This work is based in part on observations made with the Spitzer Space Telescope, which was operated by the Jet Propulsion Laboratory, California Institute of Technology under a contract with NASA. Support for this work was provided by NASA through an award issued by JPL/Caltech. This work made use of v2.2.1 of the Binary Population and Spectral Synthesis (BPASS) models as described in \citet{Eldridge2017} and \citet{Stanway2018}.

This publication makes use of data products from the Two Micron All Sky Survey, which is a joint project of the University of Massachusetts and the Infrared Processing and Analysis Center/California Institute of Technology, funded by the National Aeronautics and Space Administration and the National Science Foundation. This paper includes data gathered with the 6.5 meter Magellan Telescopes located at Las Campanas Observatory, Chile. This publication makes use of data products from the Wide-field Infrared Survey Explorer, which is a joint project of the University of California, Los Angeles, and the Jet Propulsion Laboratory/California Institute of Technology, funded by the National Aeronautics and Space Administration. This research is based on observations made with the NASA/ESA Hubble Space Telescope obtained from the Space Telescope Science Institute, which is operated by the Association of Universities for Research in Astronomy, Inc., under NASA contract NAS 5-26555. These observations are associated with program 15645. Observations were also obtained from the Hubble Legacy Archive, which is a collaboration between the Space Telescope Science Institute (STScI/NASA), the Space Telescope European Coordinating Facility (ST-ECF/ESAC/ESA) and the Canadian Astronomy Data Centre (CADC/NRC/CSA). This work made use of data from the Las Cumbres Observatory network. We thank the Las Cumbres Observatory and its staff for its continuing support of the ASAS-SN project. ASAS-SN is supported by the Gordon and Betty Moore Foundation through grant GBMF5490 to the Ohio State University, and NSF grants AST-1515927 and AST-1908570. Development of ASAS-SN has been supported by NSF grant AST-0908816, the Mt. Cuba Astronomical Foundation, the Center for Cosmology and AstroParticle Physics at the Ohio State University, the Chinese Academy of Sciences South America Center for Astronomy (CAS- SACA), and the Villum Foundation. This research has made use of the SVO Filter Profile Service\footnote{\url{http://svo2.cab.inta-csic.es/theory/fps/}} supported from the Spanish MINECO through grant AYA2017-84089. We acknowledge ESA Gaia, DPAC and the Photometric Science Alerts Team\footnote{\url{http://gsaweb.ast.cam.ac.uk/alerts}}. This project used data obtained with the Dark Energy Camera (DECam), which was constructed by the Dark Energy Survey (DES) collaborating institutions: Argonne National Lab, University of California Santa Cruz, University of Cambridge, Centro de Investigaciones Energeticas, Medioambientales y Tecnologicas-Madrid, University of Chicago, University College London, DES-Brazil consortium, University of Edinburgh, ETH-Zurich, University of Illinois at Urbana-Champaign, Institut de Ciencies de l'Espai, Institut de Fisica d'Altes Energies, Lawrence Berkeley National Lab, Ludwig-Maximilians Universitat, University of Michigan, National Optical Astronomy Observatory, University of Nottingham, Ohio State University, University of Pennsylvania, University of Portsmouth, SLAC National Lab, Stanford University, University of Sussex, and Texas A$\&$M University. Funding for DES, including DECam, has been provided by the U.S. Department of Energy, National Science Foundation, Ministry of Education and Science (Spain), Science and Technology Facilities Council (UK), Higher Education Funding Council (England), National Center for Supercomputing Applications, Kavli Institute for Cosmological Physics, Financiadora de Estudos e Projetos, Fundação Carlos Chagas Filho de Amparo a Pesquisa, Conselho Nacional de Desenvolvimento Científico e Tecnológico and the Ministério da Ciência e Tecnologia (Brazil), the German Research Foundation-sponsored cluster of excellence ``Origin and Structure of the Universe" and the DES collaborating institutions. Based on observations obtained with XMM-Newton, an ESA science mission with instruments and contributions directly funded by ESA Member States and NASA.Based on observations obtained at the international Gemini Observatory (GS-2016B-Q-22), a program of NSF’s NOIRLab, which is managed by the Association of Universities for Research in Astronomy (AURA) under a cooperative agreement with the National Science Foundation. on behalf of the Gemini Observatory partnership: the National Science Foundation (United States), National Research Council (Canada), Agencia Nacional de Investigaci\'{o}n y Desarrollo (Chile), Ministerio de Ciencia, Tecnolog\'{i}a e Innovaci\'{o}n (Argentina), Minist\'{e}rio da Ci\^{e}ncia, Tecnologia, Inova\c{c}\~{o}es e Comunica\c{c}\~{o}es (Brazil), and Korea Astronomy and Space Science Institute (Republic of Korea). This paper includes data gathered with the Nordic Optical Telescope (PI Stritzinger) at the Observatorio del Roque de los Muchachos, La Palma, Spain. We thank Norbert Schartel for rapid approval and scheduling of ToO observations (ObsID 0794580101). Based on observations collected at the European Organisation for Astronomical Research in the Southern Hemisphere under ESO programmes 1103.D-0328, 1103.D-0328, 199.D-0143, 197.D-1075, 191.D-0935 (PESSTO, ePESSTO and ePESSTO+). In addition, data taken under ESO programmes 098.A-9099, 0104.A-9099, 0100.D-0865, 098.D-0692, 100.D-0341, and 0103.D-0139(A) are presented here. Based in part on data acquired at the ANU 2.3m Telescope at Siding Spring Observatory. We acknowledge the traditional owners of the land on which the telescope stands, the Gamilaraay people, and pay our respects to elders past, present, and emerging. Part of the funding for GROND (both hardware as well as personnel) was generously granted from the Leibniz-Prize to Prof. G. Hasinger (DFG grant HA 1850/28-1). The OGLE project has received funding from the National Science Centre, Poland, grant MAESTRO 2014/14/A/ST9/00121 to AU. LCO data have been obtained via OPTICON proposals and was obtained as part of the Global Supernova Project. The OPTICON project has received funding from the European Union's Horizon 2020 research and innovation programme under grant agreement No 730890.

\section*{Data availability}

The spectroscopic data underlying this article are available in the Weizmann Interactive Supernova Data Repository at \url{https://wiserep.weizmann.ac.il/}. The photometric data underlying this article are available in the article and in its online supplementary material.



\bibliographystyle{mnras}
\bibliography{AT2016jbu_Paper_I}

\appendix

\section{Author Affiliations}\label{app:affiliations}
$^{1}$   School of Physics, O’Brien Centre for Science North, University College Dublin, Belfield, Dublin 4, Ireland\\ 
$^{2}$   The Oskar Klein Centre, Department of Physics, AlbaNova, Stockholm University, SE-106 91 Stockholm, Sweden\\ 
$^{3}$   INAF-Osservatorio Astronomico di Padova, Vicolo dell’Osservatorio 5, I-35122 Padova, Italy\\ 
$^{4}$   Department of Physics and Astronomy, University of Turku, FI-20014, Turku, Finland\\ 
$^{5}$   The Department of Physics, The University of Auckland, Private Bag 92019, Auckland, New Zealand\\ 
$^{6}$   The Oskar Klein Centre, Department of Astronomy, AlbaNova, Stockholm University, SE-106 91 Stockholm, Sweden\\ 
$^{7}$   Max-Planck-Institut f\"{u}r Extraterrestrische Physik, Giessenbachstra\ss e 1, 85748 Garching, Germany\\ 
$^{8}$   Department of Astronomy, The Ohio State University, 140 W. 18th Avenue, Columbus, OH 43210, USA\\ 
$^{9}$   Center for Cosmology and AstroParticle Physics (CCAPP), The Ohio State University, 191 W. Woodruff Avenue, Columbus, OH 43210, USA\\ 
$^{10}$ Department of Physics and Astronomy, Texas A$\&$M University, 4242 TAMU, College Station, TX 77843, USA\\ 
$^{11}$ Cerro Tololo Inter-American Observatory, NSF’s National Optical-Infrared Astronomy Research Laboratory, Casilla 603, La Serena, Chile\\ 
$^{12}$ Institut d’Astrophysique de Paris (IAP), CNRS $\&$ Sorbonne Universite, France\\ 
$^{13}$ Kavli Institute for Astronomy and Astrophysics, Peking University, Yi He Yuan Road 5, Hai Dian District, Beijing 100871, China\\ 
$^{14}$ NINAF-Osservatorio Astronomico di Padova, Vicolo dell’Osservatorio 5, I-35122 Padova, Italy\\ 
$^{15}$ Institute of Space Sciences (ICE, CSIC), Campus UAB, Carrer de Can Magrans s/n, 08193 Barcelona, Spain\\ 
$^{16}$ Steward Observatory, University of Arizona, 933 North Cherry Avenue, Tucson, AZ 85721-0065, USA\\ 
$^{17}$ Department of Physics, Florida State University, 77 Chieftan Way, Tallahassee, FL 32306, USA\\ 
$^{18}$ Tuorla Observatory, Department of Physics and Astronomy, FI-20014 University of Turku, Finland.\\ 
$^{19}$ Finnish Centre for Astronomy with ESO (FINCA), FI-20014 University of Turku, Finland\\ 
$^{20}$ CBA Kleinkaroo, Calitzdorp, South Africa\\ 
$^{21}$ Departamento de Ciencias F\'isicas, Universidad Andres Bello, Avda. Republica 252, Santiago, 8320000, Chile\\ 
$^{22}$ Millennium Institute of Astrophysics, Santiago, Chile\\ 
$^{23}$ Department of Astronomy/Steward Observatory, 933 North Cherry Avenue, Rm. N204, Tucson, AZ 85721-0065, USA\\ 
$^{24}$ Institute for Astronomy, University of Hawai’i, 2680 Woodlawn Drive, Honolulu, HI 96822, USA\\ 
$^{25}$ Astrophysics Research Centre, School of Maths and Physics, Queen’s University Belfast, Belfast BT7 1NN, UK\\ 
$^{26}$ Mt Stromlo Observatory, The Research School of Astronomy and Astrophysics, Australian National University, ACT 2601, Australia\\ 
$^{27}$ National Centre for the Public Awareness of Science, Australian National University, Canberra, ACT 2611, Australia\\ 
$^{28}$ Kavli Institute for Astronomy and Astrophysics, Peking University, Yi He Yuan Road 5, Hai Dian District, Beijing 100871, China\\ 
$^{29}$ Astronomical Observatory, University of Warsaw, Al. Ujazdowskie 4, 00-478 Warszawa, Poland\\ 
$^{30}$ Aix Marseille Univ, CNRS, CNES, LAM, Marseille, France\\ 
$^{31}$ Unidad Mixta Internacional Franco-Chilena de Astronom\'ia, CNRS/INSU UMI 3386 and Instituto de Astrof\'isica, Pontificia Universidad Cat\'olica de Chile, Santiago, Chile\\ 
$^{32}$ Institute of Astronomy, Madingley Road, Cambridge,CB3 0HA, UK\\ 
$^{33}$ RHEA Group for ESA, European Space Astronomy Centre (ESAC-ESA), Madrid, Spain\\ 
$^{34}$ Institute of Space Sciences (ICE, CSIC), Campus UAB, Carrer de Can Magrans, s/n, E-08193 Barcelona, Spain\\ 
$^{35}$ Tuorla Observatory, Department of Physics and Astronomy, FI-20014 University of Turku, Finland\\ 
$^{36}$ Kavli Institute for Cosmology, Institute of Astronomy, Madingley Road, Cambridge, CB3 0HA, UK\\ 
$^{37}$ Las Cumbres Observatory, 6740 Cortona Drive, Suite 102, Goleta, CA 93117-5575, USA\\ 
$^{38}$ Department of Physics, University of California, Santa Barbara, CA 93106-9530, USA\\ 
$^{39}$ The Observatories of the Carnegie Institution for Science, 813 Santa Barbara St., Pasadena, CA 91101, USA\\ 
$^{40}$ School of Physics $\&$ Astronomy, Cardiff University, Queens Buildings, The Parade, Cardiff, CF24 3AA, UK\\ 
$^{41}$ School of Physics and Astronomy, University of Southampton, Southampton, Hampshire, SO17 1BJ, UK\\ 
$^{42}$ School of Physics, Trinity College Dublin, The University of Dublin, Dublin 2, Ireland\\ 
$^{43}$ Department of Physics, University of the Free State, PO Box 339, Bloemfontein 9300, South Africa\\ 
$^{44}$ Carnegie Observatories, Las Campanas Observatory, Colina El Pino, Casilla 601, Chile\\ 
$^{45}$ Birmingham Institute for Gravitational Wave Astronomy and School of Physics and Astronomy, University of Birmingham, Birmingham B15 2TT, UK\\ 
$^{46}$ Institute for Astronomy, University of Edinburgh, Royal Observatory, Blackford Hill, EH9 3HJ, UK\\ 
$^{47}$ Nucleo de Astronomıa de la Facultad de Ingenierıa y Ciencias, Universidad Diego Portales, Av. Ej´ercito 441, Santiago, Chile\\ 
$^{48}$ Department of Physics and Astronomy University of North Carolina at Chapel Hill Chapel Hill, NC 27599, USA\\ 
$^{49}$ Department of Physics, Florida State University, 77 Chieftain Way, Tallahassee, FL 32306-4350, USA\\ 
$^{50}$ Department of Physics and Astronomy, Aarhus University, Ny Munkegade, DK-8000 Aarhus C, Denmark\\ 
$^{51}$ Department of Physics, University of Warwick, Coventry CV4 7 AL, UK\\ 
$^{52}$ Astrophysics Research Centre, School of Mathematics and Physics, Queen`s University Belfast, Belfast BT7 1NN, UK\\ 
$^{53}$ Institute of Space Sciences (ICE, CSIC), Campus UAB, Carrer de Can Magrans, s/n, E-08193 Barcelona, Spain.
$^{54}$ Center for Astrophysics \textbar{} Harvard \& Smithsonian, 60 Garden Street, Cambridge, MA 02138-1516, USA
$^{55}$ The NSF AI Institute for Artificial Intelligence and Fundamental Interactions

\onecolumn
\section{Photometry Tables}\label{app:appendix_phot}
\begin{longtable}{lrlllllllllllllll}
\caption{Sample of full photometry table for \jbu. All measurements were carried out using {\sc AutoPhoT}. Phase is with respect to $V$-band maximum of {\it Event B}. Limiting magnitudes listed where \jbu could not be detected, and 1$\sigma$ errors are given in parentheses. {\it UBVRIJHK} filters are in Vegamags, {\it ugriz} are in AB magnitudes.
    Full photometry table available online.
                }\\
\toprule
       Date &      MJD &  Phase (d) &  u &  g &  r &  i &  z &  U &         B &         V &         R &  I &  J &  H &  K &   Instrument \\
\midrule
\endhead
\midrule
\multicolumn{17}{r}{{Continued on next page}} \\
\midrule
\endfoot

\bottomrule
\endlastfoot
 1999-12-26 &  51538.5 &  $-$6245.9 &  - &  - &  - &  - &  - &  - &  $>$22.63 &         - &         - &  - &  - &  - &  - &          WFI \\
 2000-02-17 &  51591.0 &  $-$6193.4 &  - &  - &  - &  - &  - &  - &  $>$22.66 &  $>$21.94 &  $>$22.80 &  - &  - &  - &  - &          WFI \\
 2000-04-05 &  51639.0 &  $-$6145.4 &  - &  - &  - &  - &  - &  - &         - &         - &  $>$23.33 &  - &  - &  - &  - &          WFI \\
 2001-02-04 &  51944.0 &  $-$5840.4 &  - &  - &  - &  - &  - &  - &         - &  $>$22.37 &  $>$23.20 &  - &  - &  - &  - &          WFI \\
 2005-03-13 &  53442.0 &  $-$4342.4 &  - &  - &  - &  - &  - &  - &         - &  $>$22.54 &         - &  - &  - &  - &  - &  CTIO+MOSAIC \\
 2005-03-14 &  53443.0 &  $-$4341.4 &  - &  - &  - &  - &  - &  - &         - &  $>$22.59 &  $>$20.50 &  - &  - &  - &  - &  CTIO+MOSAIC \\
 2006-01-29 &  53764.0 &  $-$4020.4 &  - &  - &  - &  - &  - &  - &         - &  $>$23.19 &         - &  - &  - &  - &  - &          WFI \\
 2006-01-29 &  53764.5 &  $-$4019.9 &  - &  - &  - &  - &  - &  - &         - &  $>$23.23 &         - &  - &  - &  - &  - &          WFI \\
 2006-01-30 &  53765.0 &  $-$4019.4 &  - &  - &  - &  - &  - &  - &  $>$24.36 &         - &         - &  - &  - &  - &  - &          WFI \\
 2006-10-06 &  54014.0 &  $-$3770.4 &  - &  - &  - &  - &  - &  - &         - &         - &  $>$16.36 &  - &  - &  - &  - &       Prompt \\
\end{longtable}


\begin{table*}
\setlength{\tabcolsep}{4pt} 
\caption{Properties of \ip-like transient events. Values reported are used consistently throughout this work. The time of peak is with respect to the \textit{Event B} maximum. Where quoted, $^{56}$Ni masses are upper limits.}
\label{tab:objects}
\begin{threeparttable}
\begin{tabular}{llllllll}
\hline
Transient & z & A$_V$ [mag] $^1$ & $\mu$ [mag] & Peak (MJD) $^2$ & $^{56}Ni$~[\msun] & Reference \\
\hline
\jbu & 0.00489 & 0.556 & 31.60 & 57784 & $\le$0.016 & This paper; \paperII; \citet{Cartier17}\\
\ip & 0.00572 & 0.054 & 31.55 & 56203 & $\le$0.020 & \citet{Fraser13,Pastorello2013} \\
\gc & 0.00340 & 1.253 & 30.46 & 56544 & $\le$0.004 & \citet{Reguitti2018} \\
\bh & 0.00644 & 0.062 & 32.40 & 57166 & $\le$0.003 & \citet{Thone2017,Elisa-Rosa2016}\\
\bdu & 0.0173 & 0.041 & 34.37 & 57541 & -  & \citet{Pastorello_2017} \\
\lsq & 0.029 & 0.052 & 35.43 & 56406 & -  & \citet{Tartaglia2016} \\
\al & 0.0065 & 0.032 & 31.80 & 50265 & -  & \citet{Benetti_2016} \\
\cnf & 0.02376 & 0.118 & 34.99 & 58293 & -  & \citet{Pastorello2019} \\
\hline
\end{tabular}
\begin{tablenotes}\footnotesize
\item[1] Galactic Extinction only. If $A_V$ not mentioned in reference, we take values from NED.
\item[2] With respect to \textit{Event B} maximum light in \textit{V}-band.
\end{tablenotes}
\end{threeparttable}
\end{table*}

\bsp	
\label{lastpage}
\end{document}